\documentclass[reprint,superscriptaddress,preprintnumbers,amsmath,amssymb,aps,prb]{revtex4-2}
\usepackage[english]{babel}
\usepackage{graphicx}
\usepackage{dcolumn}
\usepackage{bm}
\usepackage{xcolor}
\usepackage{braket}
\usepackage[unicode=true,colorlinks=true,citecolor=blue]{hyperref}
\usepackage[normalem]{ulem}

\begin{document}


\title{Trion magnetic polarons in (Cd,Mn)Te/(Cd,Mn,Mg)Te quantum wells}


\author{F. Godejohann}
 \email{felix.godejohann@tu-dortmund.de}
\affiliation{Experimentelle Physik 2, Technische Universität Dortmund, 44227 Dortmund, Germany}

\author{R. R. Akhmadullin}
\affiliation{Ioffe Institute, Russian Academy of Science, 194021 St. Petersburg, Russia}

\author{K. V. Kavokin}
\affiliation{Spin Optics Laboratory, St. Petersburg State University, 198504 St. Petersburg, Russia}

\author{D. R. Yakovlev}
\affiliation{Experimentelle Physik 2, Technische Universität Dortmund, 44227 Dortmund, Germany}
\affiliation{Ioffe Institute, Russian Academy of Science, 194021 St. Petersburg, Russia}

\author{I. A. Akimov}
\affiliation{Experimentelle Physik 2, Technische Universität Dortmund, 44227 Dortmund, Germany}
\affiliation{Ioffe Institute, Russian Academy of Science, 194021 St. Petersburg, Russia}

\author{B. R. Namozov}
\affiliation{Ioffe Institute, Russian Academy of Science, 194021 St. Petersburg, Russia}

\author{Yu. G. Kusrayev}
\affiliation{Ioffe Institute, Russian Academy of Science, 194021 St. Petersburg, Russia}

\author{G. Karczewski}
\affiliation{Institute of Physics, Polish Academy of Sciences, 02668 Warsaw, Poland}

\author{T. Wojtowicz}
\affiliation{International Research Centre MagTop, Institute of Physics, Polish Academy of Sciences, 02668 Warsaw, Poland}

\author{M. Bayer}
\affiliation{Experimentelle Physik 2, Technische Universität Dortmund, 44227 Dortmund, Germany}
\affiliation{Ioffe Institute, Russian Academy of Science, 194021 St. Petersburg, Russia}

\date{\today}

\begin{abstract}
A trion magnetic polaron formed by the exchange interaction of a positively charged exciton (trion) with localized spins of Mn$^{2+}$ ions is found experimentally in a 4\,nm wide Cd$_{0.98}$Mn$_{0.02}$Te/Cd$_{0.78}$Mn$_{0.02}$Mg$_{0.2}$Te quantum well containing resident holes. The experiment is performed at a temperature of 1.6\,K using resonant excitation of the trion with circularly polarized light. The trion is formed from a resident hole, which is in a hole magnetic polaron state, and a photogenerated electron-hole pair. The dynamical evolution from the hole magnetic polaron to the trion magnetic polaron is accompanied by a spin-flip of the electron, which results in negative circular polarization of the photoluminescence. The degree of circular polarization reaches $-8\%$ at zero magnetic field and strongly decreases in transverse magnetic fields exceeding 0.2~T.  Our model considerations show that different localization sizes of the resident and photogenerated holes and the resulting difference in their exchange interaction with the Mn$^{2+}$ spins maintains Mn spin polarization.  The resulting exchange field of Mn acting on the electron provides a robust spin polarization of the trion magnetic polaron.  We evaluate the electron exchange energy in the T$^+$MP to be 0.19~meV, and the T$^+$MP binding energy to be about 0.5 - 1\,meV.
\end{abstract}

\maketitle


\section{\label{sec:introduction}Introduction}

Spin orientation and control in semiconductors are important for modern quantum information technologies \cite{awschalom_semiconductor_2002, zutic_spintronics_2004}. A convenient method to generate a carrier spin polarization in semiconductors is provided by optical orientation (OO) using circularly polarized light. In diluted magnetic semiconductors (DMS), photogenerated charge carriers are coupled to the localized spins of magnetic ions by the strong $sp$-$d$ exchange interaction. The most common magnetic dopant for II-VI semiconductors, such as CdTe, and III-V semiconductors is manganese \cite{warnock_localized_1985, dietl_dilute_2014}. 

In general, the OO in DMS semiconductors is quite small as the photogenerated spin polarization of charge carriers experiences an efficient, fast spin relaxation by means of the strong exchange interaction with localized Mn$^{2+}$ spins \cite{crooker_terahertz_1996, crooker_optical_1997, akimoto_larmor_1998}. However, the spin complexes known as magnetic polarons, which are formed by a charge carrier and an ensemble of localized spins within the carrier localization volume, can keep their spin orientation for long times. There are several types of magnetic polarons, e.g., donor-bound and acceptor-bound magnetic polarons in bulk \cite{wolff_theory_1988, zhukov_optical_2019} or localized magnetic polarons in low dimensional structures such as quantum wells (QWs) \cite{akimov_dynamics_2017} and quantum dots \cite{seufert_dynamical_2001}. Besides exciton magnetic polarons (EMP), which are bound complexes consisting of a photogenerated exciton and about a hundred Mn$^{2+}$ spins, a recent study has shown the presence of resident hole magnetic polarons in (Cd,Mn)Te-based QWs \cite{zhukov_optical_2016}. Hole magnetic polarons (HMP) are prospective candidates for achieving a long-lived spin polarization. Optical accessibility of the spin-polarized hole magnetic polaron can become an important technique to utilize such states. 

In this paper we report on the optical readout of spin-polarized hole magnetic polarons via selectively excited trion magnetic polarons (T$^+$MPs) in (Cd,Mn)Te/(Cd,Mn,Mg)Te QWs. The T$^+$MP is a bound states of a photogenerated electron-hole pair and a resident HMP. Due to the presence of two holes and only one electron, the T$^+$MP can be understood as a trion (T$^+$) in a magnetic polaron state with manganese. In case of selective excitation of the T$^+$ state, we find a sign inversion of the OO with respect to nonselective excitation. Its absolute value significantly changes from about $+2$\,\% to $-8$\,\% within an excitation energy range of several meV only. We attribute the change in the OO to an electron spin flip process during the T$^+$MP formation and optical readout of spin-polarized resident HMPs. Moreover, from depolarization of the PL in an external transverse magnetic field we evaluate the hole exchange field acting on the Mn ions for the T$^+$MP complex.

The paper is organized as follows. 
In Sec.~\ref{sec:sample_experiment} the sample and experimental techniques are described. 
In Sec.~\ref{sec:experimental_results} the experimental data of time-integrated photoluminescence (TIPL) and time-resolved photoluminescence (TRPL) measurements are presented. 
The section focuses on the experimental findings related to the sign inversion of the OO depending on the photon energy.
Moreover, the dependence of the OO on an applied transverse magnetic field is studied experimentally. 
In Sec.~\ref{sec:model_OO} we consider the scenario of T$^+$MP formation, which yields sign inversion of the OO under selective excitation of the trion states. 
Theoretical consideration of the T$^+$MPs is given in Secs.~\ref{sec:theory} and \ref{sec:mod}.
In Sec.~\ref{sec:model_Depol} we consider the depolarization in a transverse magnetic field. In Sec.~\ref{sec:discussion} we present model calculations of the T$^+$MP characteristic energies. In conclusion, we provide an overview of different types of magnetic polarons.

\section{\label{sec:sample_experiment}Sample and experimental technique}

\begin{figure}[t]
\includegraphics[width=0.47\textwidth]{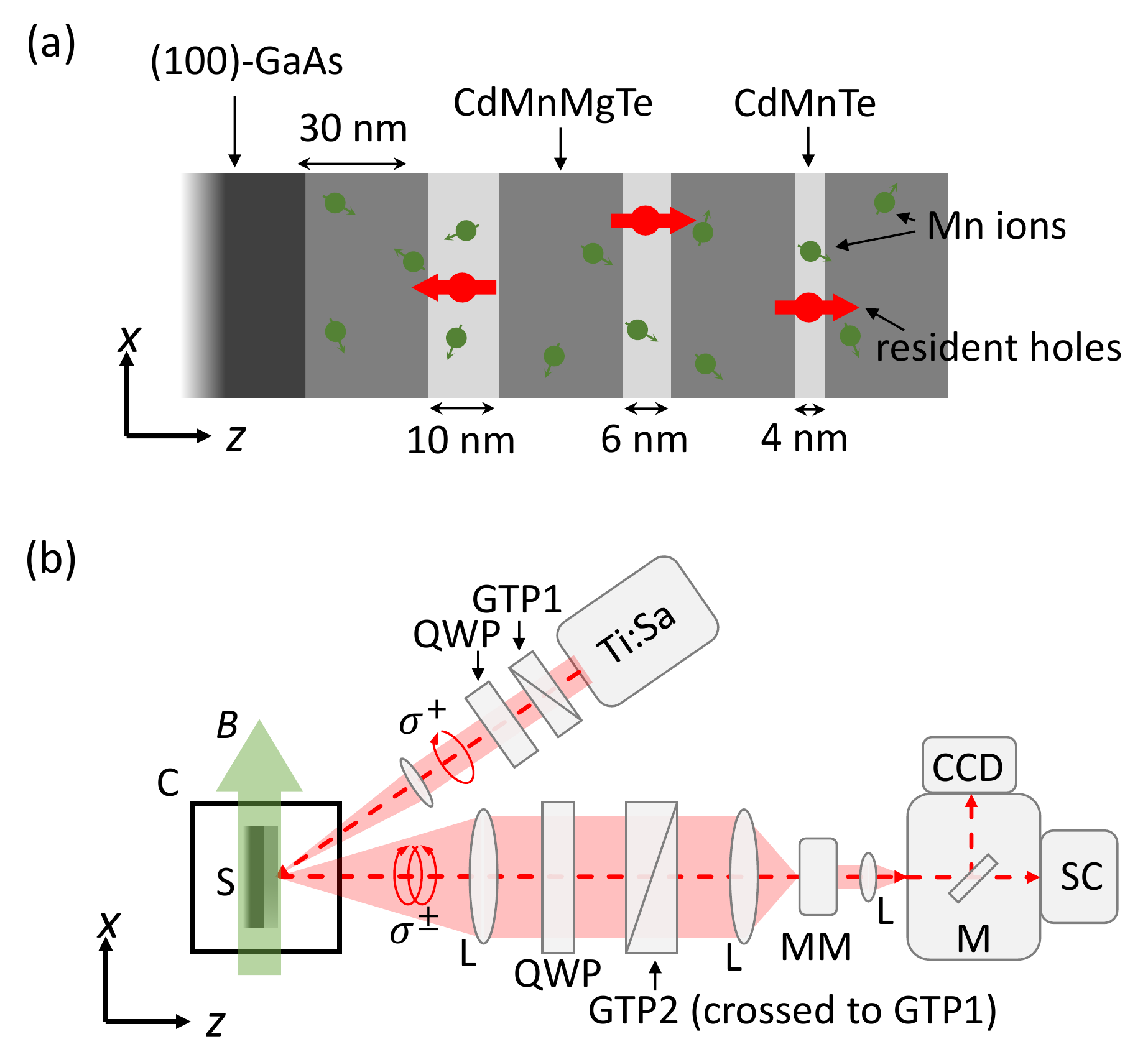}
\includegraphics[width=0.48\textwidth]{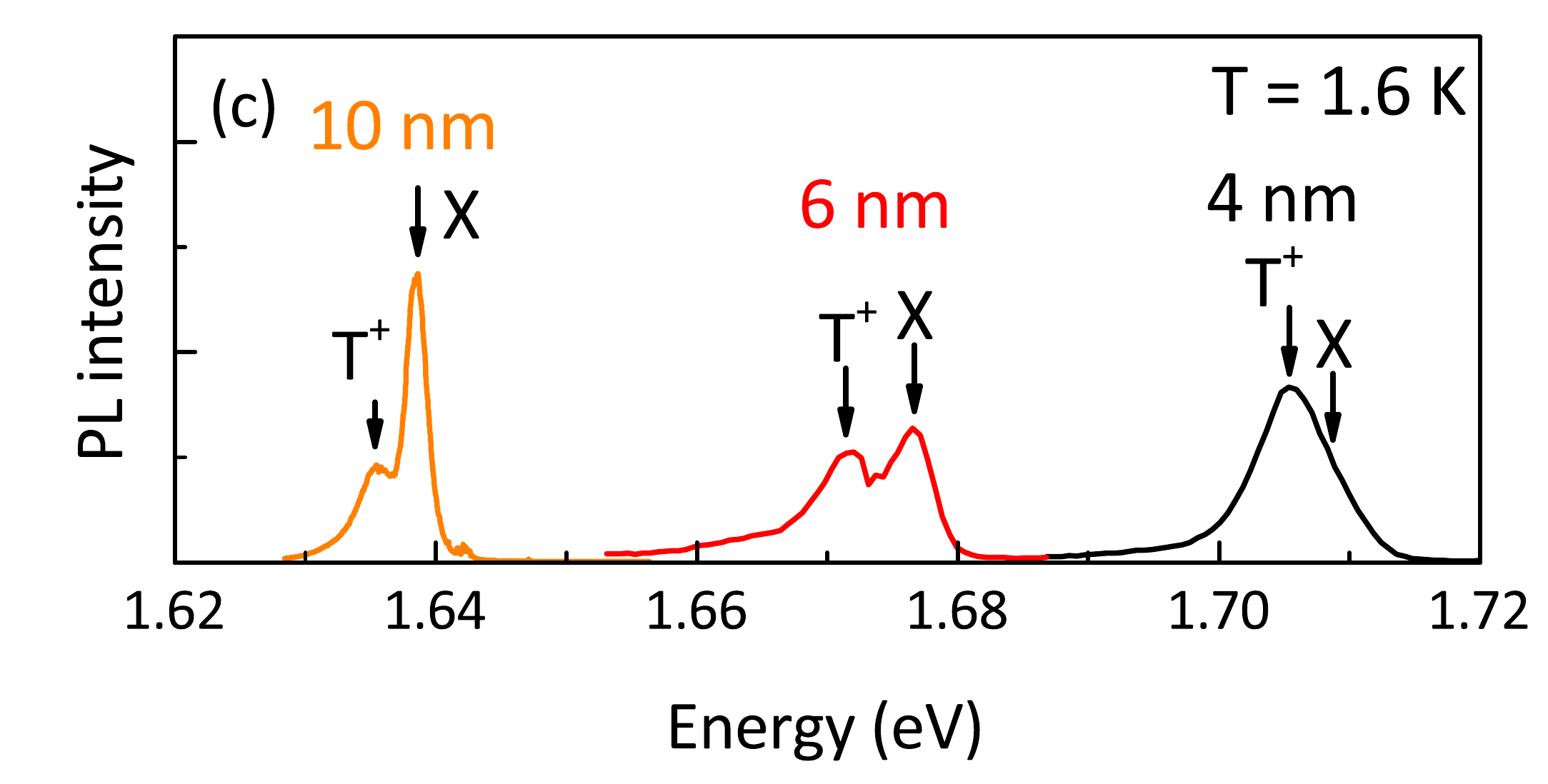}
\caption{\label{Sample_Experiment} (a) Diluted magnetic semiconductor heterostructure consisting of Cd$_{0.98}$Mn$_{0.02}$Te quantum QWs with thicknesses of 4, 6, and $10\,$nm, which are separated by 30-nm Cd$_{0.78}$Mn$_{0.02}$Mg$_{0.2}$Te barriers. The structure was grown on a bulk GaAs substrate. The magnetic Mn$^{2+}$ ions are shown as green dots and the resident holes as red dots. (b) Schematic of the experimental setup. A Ti:Sapphire (Ti:Sa) laser is used for excitation of the sample (S), which is placed in a cryostat (C). The emitted circularly polarized photoluminescence is collected by a lens (L) and analysed by a combination of a quarter-wave plate (QWP) and a Glan-Thompson prism (GTP). The laser stray light is filtered out by means of a mini-monochromator (MM). After transmission, the PL light is finally focused onto a monochromator (M) to which a charge coupled device (CCD) and a streak camera (SC) are attachjed for detection. (c) Time-integrated photoluminescence (TIPL) spectrum across the spectral range of the 4, 6, and $10\,$nm QW emission measured at $B=0\,$T for $T = 1.6\,$K using off-resonant cw excitation with a He-Ne laser, emitting photons with an energy of 1.96\,eV.}
\end{figure}

The investigated sample (041300B) contains three Cd$_{0.98}$Mn$_{0.02}$Te QWs with a low Mn$^{2+}$ concentration of 2\,\% and different thicknesses of 4, 6, and 10~nm, see Fig.~\ref{Sample_Experiment}(a). The Cd$_{0.98}$Mn$_{0.02}$Te QWs are separated by 30 nm-thick barriers of Cd$_{0.78}$Mn$_{0.02}$Mg$_{0.2}$Te, so that they can be considered as electronically decoupled. The structure was grown by molecular-beam epitaxy on a (100)-oriented GaAs substrate. The structure is nominally undoped, but the QWs contain low concentrations of resident holes \cite{zhukov_optical_2016}. 

Figure~\ref{Sample_Experiment}(a) gives a sketch of the studied structure, where the Mn$^{2+}$ ions are indicated in green and the resident holes in red. As  shown in Fig.~\ref{Sample_Experiment}(b), the sample (S) is placed in a helium bath cryostat with a split-coil superconducting solenoid, where the structure growth axis is directed along the $z$-axis. The experiments are performed at cryogenic temperatures varying from 1.6~K up to 15~K.  Magnetic fields up to 1~T can be applied perpendicular to the $z$-axis (Voigt geometry).

The experimental technique is based on conventional photoluminescence (PL) spectroscopy, with the experimental scheme shown in Fig.~\ref{Sample_Experiment}(b). 
A continuous wave (cw) or a pulsed Ti:Sapphire (Ti:Sa) laser with tunable photon energy are used for excitation of spin-oriented electron hole pairs in the studied QWs. In the pulsed mode, the repetition rate (repetition period) and duration of the laser pulses are 76.62\,MHz (13 ns) and $150\,$fs, respectively. Here, a pulse shaper is used in order to reduce the laser's spectral width of $20\,$meV to about $1\,$meV (not shown in the sketch). This enables the possibility of quasi-resonant excitation close to the exciton and trion resonances. 
For excitation of the sample, the linearly polarized laser emission is sent through a Glan-Thompson prism and a quarter-wave plate, leading to circularly polarized light, which is then focused onto the sample via a lens resulting in a laser spot with a diameter of about $100-200$~$\mu$m. The excited photoluminescence is collected by a lens and analyzed by a quarter-wave plate in combination with a Glan-Thompson prism. Another lens is used to focus the PL onto a mini monochromator, used to block scattered laser light.
Then, the PL is passed through a single stage monochromator (300\,grooves/mm grating), which is equipped with a liquid-nitrogen cooled charge coupled device (CCD) and a streak camera. With flipping a mirror inside the monochromator, it is possible to switch between time-integrated detection using the CCD and time-resolved (TR) detection using the streak camera. The overall temporal resolution of the setup is about 10~ps. For smoothing the measured curves in this paper, five data points are averaged to a single point. Under cw excitation the laser spectral width is about 0.1~$\mu$eV and a similar excitation scheme is used. For detection of the PL polarization degree an electro-optical modulator in combination with a Glan-Thompson prism is used. The PL signal is dispersed with a triple monochromator and detected with a single channel photomultiplier.

Figure~\ref{Sample_Experiment}(c) shows TIPL spectra of all three QWs for off-resonant cw excitation with the He-Ne laser (photon energy $E_\text{exc} = 1.96\,$eV). The red and orange spectra of the 6 and 10\,nm wide QWs show PL line doublets, which can be identified as emissions from exciton (X) and trion (T$^+$) states, respectively \cite{zhukov_optical_2016}. The exciton and trion energies in the 10\,nm QW read 1.639 and 1.635\,eV, respectively.
The exciton and trion energies in the 6\,nm QW read 1.677 and 1.672\,eV. The black spectrum of the 4\,nm QW demonstrates only a single line at 1.705\,eV. 
Due to inhomogeneous broadening of the exciton and trion states, they cannot be spectrally resolved \cite{zhukov_optical_2016}.

\section{\label{sec:experimental_results}Experimental results}

\begin{figure}[b]
\includegraphics[width=0.48\textwidth]{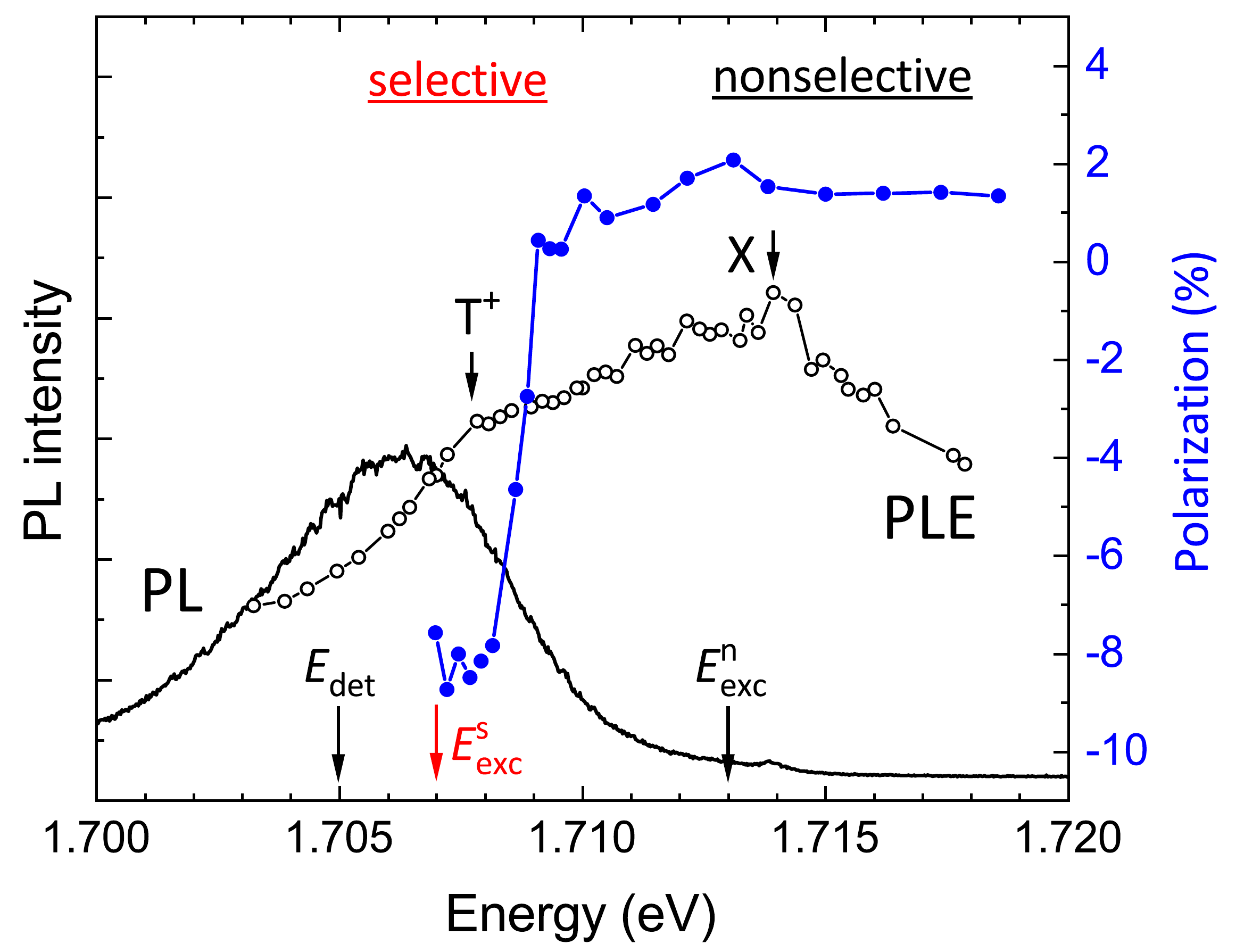}
\caption{\label{Exc_energy_dep_OO} Close-up of time-integrated photoluminescence (TIPL) spectra of the $4\,$nm QW at $T = 1.6\,$K under cw excitation with the He-Ne laser (photon energy 1.96 eV). Open circles represent the PLE spectrum taken under cw excitation and detected at $1.702\,$eV in order to clearly observe the T$^+$ excitation resonance at 1.708\,meV. The OO is represented by blue solid circles and taken at $E_\text{det} = 1.705\,$eV for different cw excitation energies.} 
\end{figure}

In the following, we focus on the OO of the PL emitted by the $4\,$nm QW (PL spectrum in Fig.~\ref{Exc_energy_dep_OO}). The OO is characterized by the degree of circular polarization with respect to the polarization of excitation. For $\sigma^+$ excitation it is given by
\begin{align}
P_\text{oo}
	&= \frac{I_+ - I_-}{I_+ + I_-},
	\label{eq:Pol}
\end{align}
where $I_+$ and $I_-$ are the emission intensities detected in $\sigma^+$ and $\sigma^-$ polarization, respectively. A positive OO corresponds to the case when the polarization of the PL has the same sign as the polarization of the excitation light; i.e. $P_\text{oo} > 0$. The excitation light is always taken to be $\sigma^+$ polarized with an adjustable photon energy $E_\text{exc}$. The average power is about $5$~mW.  As shown in Fig.~\ref{Exc_energy_dep_OO} by the blue circles, the measured OO at $E_\text{det} = 1.705\,$eV strongly depends on the excitation energy $E_\text{exc}$, which is varied from 1.707 to 1.718~eV. The OO has two regimes: 

(I) For $E_\text{exc} > 1.709$~eV, $P_\text{oo} > 0$ and takes on values of about $2$\%. 

(II) For $E_\text{exc} < 1.709$~eV, $P_\text{oo} < 0$ and has values of about $-8$\%. 

Note, that to the best of our knowledge a sign inversion of the OO for selective optical excitation in II-VI DMS heterostructures has not been reported so far. Understanding the physical origin of this unusual regime is the main motivation of our study.

In order to identify the exciton state, which is responsible for the negative OO, a photoluminescence excition (PLE) spectrum was measured with PL detection at $1.702$~eV. It is shown with open circles in Fig.~\ref{Exc_energy_dep_OO}. In spite the strong inhomogeneous broadening, it is possible to distinguish the spectral positions of excitons and trions around 1.714 and 1.708\,eV, respectively. 
Hence, the cases (I) and (II) can be distinguished according to nonselective and selective excitation of the T$^+$ state. 
The positively charged exciton, T$^+$, is formed from a resident hole localized in the QW and a photogenerated exciton \cite{zhukov_optical_2016}. 


For the sake of simplicity, this study focuses on the experimental results obtained for the $4\,$nm QW. 
While the observed behaviour is qualitatively similar in all QWs, the 4$\,$nm QW provides the strongest effects compared to the 6 and 10\,nm QWs.

\begin{figure}
\includegraphics[width=0.48\textwidth]{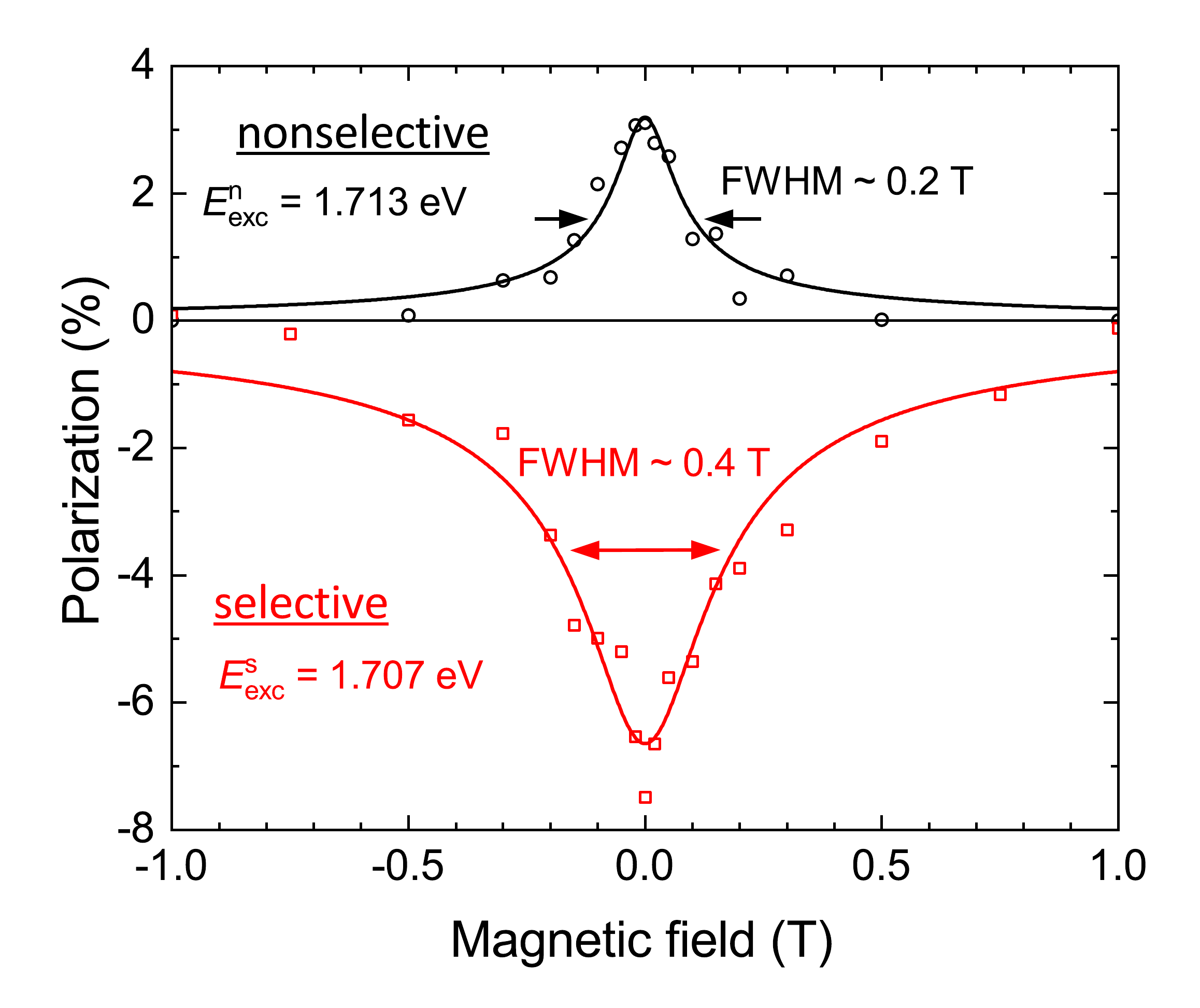}
\caption{\label{Polarization_Voigt} Depolarization curves of the OO in the nonselective and the selective regime using cw excitation with photon energies of $E^\text{n}_\text{exc} = 1.713\,$eV and $E^\text{s}_\text{exc} = 1.707\,$eV, respectively, at $T=1.6\,$K. The fits are done according to Eq.~(\ref{eq:Pol_Bx_0}) and discussed in Sec.~\ref{sec:model_Depol}.}
\end{figure}

In ~\ref{Polarization_Voigt} we show the dependence of the OO on a transverse magnetic field $\boldsymbol{B} = (B,0,0)^T$ oriented along the $x$-direction, normal to the optical axis (Voigt geometry). It was measured for two representative cases for nonselective and selective excitation.  The used excitation energies $E^\text{n}_\text{exc}$ and $E^\text{s}_\text{exc}$ are indicated in Fig.~\ref{Exc_energy_dep_OO} by the black and red arrows, respectively. Figure~\ref{Polarization_Voigt} shows the experimental findings in form of depolarization curves, where an increasing magnetic field suppresses the OO until it vanishes. In case of nonselective excitation at $E^\text{n}_\text{exc} = 1.713\,$eV, the OO is positive and possesses a full width at half maximum (FWHM) of about 0.2$\,$T. A similar depolarization curve is observed in case of selective excitation energy at $E^\text{s}_\text{exc} = 1.707\,$eV, however, it is negative. The FWHM of the red depolarization curve is about $0.4\,$T being almost twice larger than the one for nonselective excitation. 

At first glance, the shapes of the presented magnetic field induced depolarization curves in Fig.~\ref{Polarization_Voigt} suggest as mechanism the Hanle effect for the OO of photogenerated carriers, which is well-known to be present in (non-magnetic) CdTe/(Cd,Mg)Te QWs \cite{zhukov_spin_2007, astakhov_exciton_2007}. In this study, however, we show that the presented depolarization curves have a different origin than the conventional Hanle effect that is due to the Larmor precession of carrier spins. In order to further investigate the mechanism of depolarization, TRPL measurements are performed. 

\begin{figure}
\includegraphics[width=0.48\textwidth]{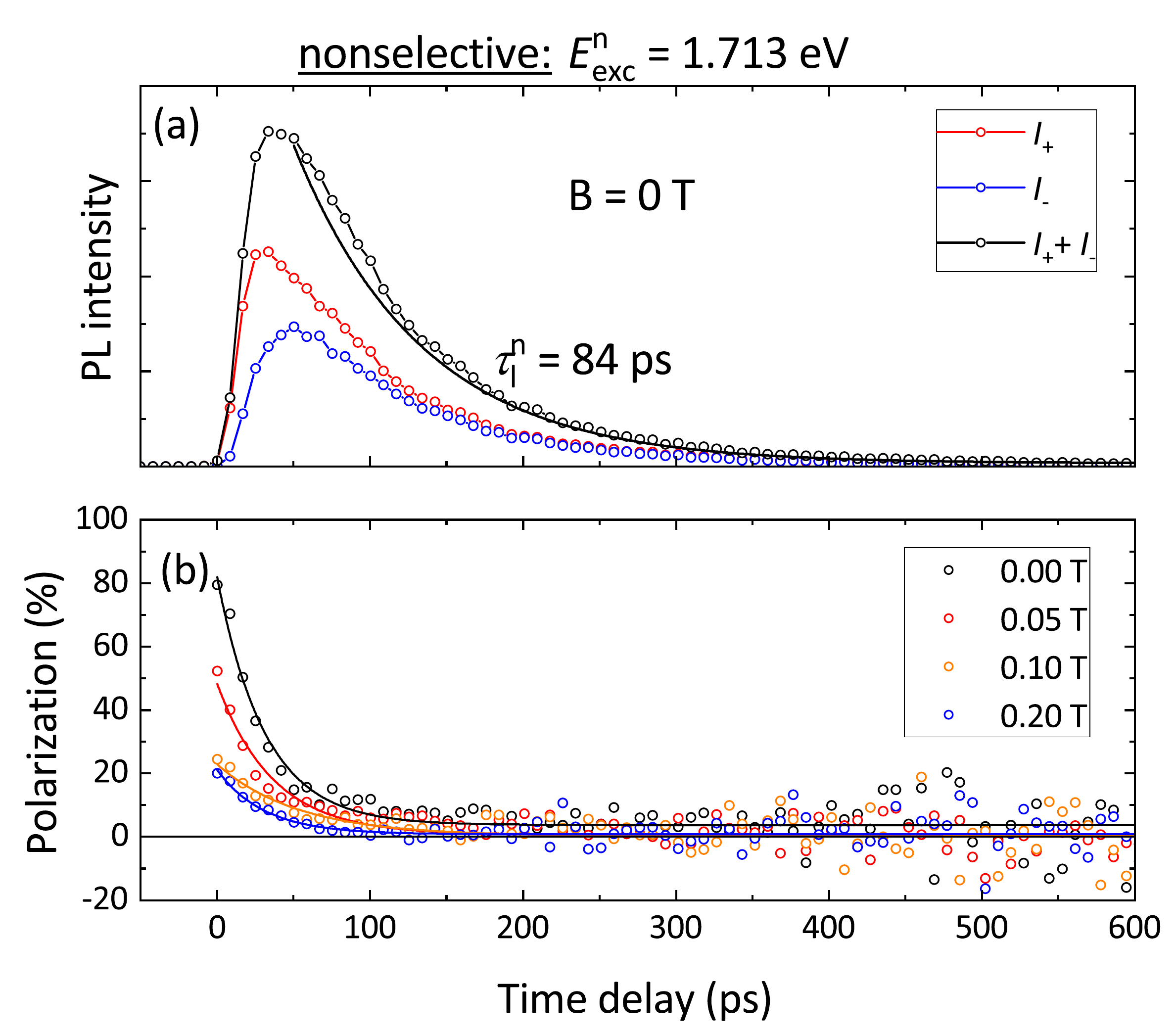}
\caption{\label{TRPL_RegimeI}  (a) Time-resolved photoluminescence (TRPL) measurements using $\sigma^+$ polarized excitation at the nonselective excitation energy $E^\text{n}_\text{exc} = 1.713\,$eV for $T=1.6\,$K. The detection energy is $E_\text{det} = 1.705\,$eV. The evaluated exciton/trion lifetime corresponding to the exponential fit reads $\tau^\text{n}_\text{l} = 84 \pm 1$ ps. (b) Time-resolved polarization measurements at various magnetic fields. The colored lines are fits using Eq.~(\ref{eq:Time_dep_OO}). The parameters of all regressions are listed in Table~\ref{tab:Positive_Regime}.}
\end{figure}

Figure~\ref{TRPL_RegimeI} shows time-resolved measurements of the PL and the (positive) OO in case of $\sigma^+$ non-selective excitation at $E^\text{n}_\text{exc} = 1.713\,$eV. Note that, due to the use of short pulses and corresponding large power densities per pulse, the average power for the TRPL measurements is about $0.5\,$mW and, therefore, smaller than in Fig.~\ref{Polarization_Voigt}. 

Figure~\ref{TRPL_RegimeI}(a) shows transient PL intensities detected in $\sigma^+$ ($I_+$) and $\sigma^-$ ($I_-$) polarized light. 
The exciton lifetime $\tau_\text{l}^\text{n}$ for nonselective excitation is evaluated using the total PL intensity, $I_+ + I_- \propto \exp{\left(-t/\tau_\text{l}\right)}$, and is given by $\tau_\text{l}^\text{n} = 84\,$ps, which is in agreement with previously reported values for localized excitons obtained in Cd$_{1-x}$Mn$_x$Te-based structures \cite{zhukov_optical_2016, mackh_localized_1994, gaj_introduction_2010}. 
The corresponding fit is shown as the black line. 

Figure~\ref{TRPL_RegimeI}(b) shows transients of OO measurements according to Eq.~(\ref{eq:Pol}). 
We observe positively valued, single exponentially decaying curves for the OO obeying the form 
\begin{align}
P_\text{oo} (t) = P_0 \, \exp{\left(-t/\tau_\text{s}\right)} + P_\infty,
\label{eq:Time_dep_OO}
\end{align}
where $P_0$ is the OO amplitude, $\tau_\text{s}$ is the spin relaxation time and $P_\infty$ is the OO offset for long times. 
The parameters evaluated from the dynamics are presented in Table~\ref{tab:Positive_Regime}. 
At $B = 0\,$T, we obtain a large OO amplitude of $P^\text{n}_0 (B = 0\,\text{T}) \approx 80\,\%$. 
This large value corresponds to the initial spin polarization of the photoexcited carriers. 
Since the spin relaxation time ($\tau^\text{n}_\text{s} = 32\,$ps) is not much shorter than the carrier lifetime ($\tau^\text{n}_\text{l} = 84\,$ps), it is expected to observe a non-zero OO in TIPL measurements, as in Fig.~\ref{Exc_energy_dep_OO}. 

\begin{table}
\begin{tabular}{|c|c|c|c||c|}
\hline
$B$ (T) & $P^\text{n}_0$ (\%)   & $\tau^\text{n}_\text{s}$ (ps) & $P^\text{n}_\infty$ (\%) & $\tau^\text{n}_\text{l}$ (ps)\\ \hline
0.00     & 78 $\pm$ 6 & 32 $\pm$ 4  & 4 $\pm$ 1   	& 84 $\pm$ 1\\ 
0.05  & 48 $\pm$ 4 & 36 $\pm$ 6  & 1 $\pm$ 1   		& 95 $\pm$ 2\\ 
0.10   & 23 $\pm$ 4 & 53 $\pm$ 17  & 0 $\pm$ 1 		& 98 $\pm$ 3\\
0.20   & 20 $\pm$ 6 & 31 $\pm$ 15  & 1 $\pm$ 1 		& 93 $\pm$ 2\\ \hline
\end{tabular}
\caption{Parameters evaluated from the fits shown in Fig.~\ref{TRPL_RegimeI}.}
\label{tab:Positive_Regime}
\end{table}

By increasing the magnetic field strength, we observe depolarization of the OO amplitude $P_0(B)$, e.g. see data at zero delay time in Fig.~\ref{TRPL_RegimeI}, similar to the depolarization observed in TIPL measurements in Fig.~\ref{Polarization_Voigt}. The obtained exciton lifetime $\tau^\text{n}_{\text{l}}$ and the spin relaxation time $\tau^\text{n}_{\text{s}}$, however, remain the same. The absence of any spin precession in the measured signals and the depolarization of $P_0$ are first hints that the physical origin of the depolarization curve in Fig.~\ref{Polarization_Voigt} is not given by the Hanle effect. 
This means that the width of the depolarization curve in Fig.~\ref{Polarization_Voigt} is not a measure for the spin and recombination dynamics, i.e. no Larmor precession is involved.

Since the noise of the OO data significantly increases with increasing time delay, we assume that the evaluated OO offsets $P^\text{n}_\infty $ in Table~\ref{tab:Positive_Regime} are close to zero. The next step is to perform a similar analysis for selective excitation at $E^\text{s}_\text{exc} = 1.707\,$eV, where the OO presented in Figs.~\ref{Exc_energy_dep_OO} and \ref{Polarization_Voigt} have negative values. 

\begin{figure}
\includegraphics[width=0.48\textwidth]{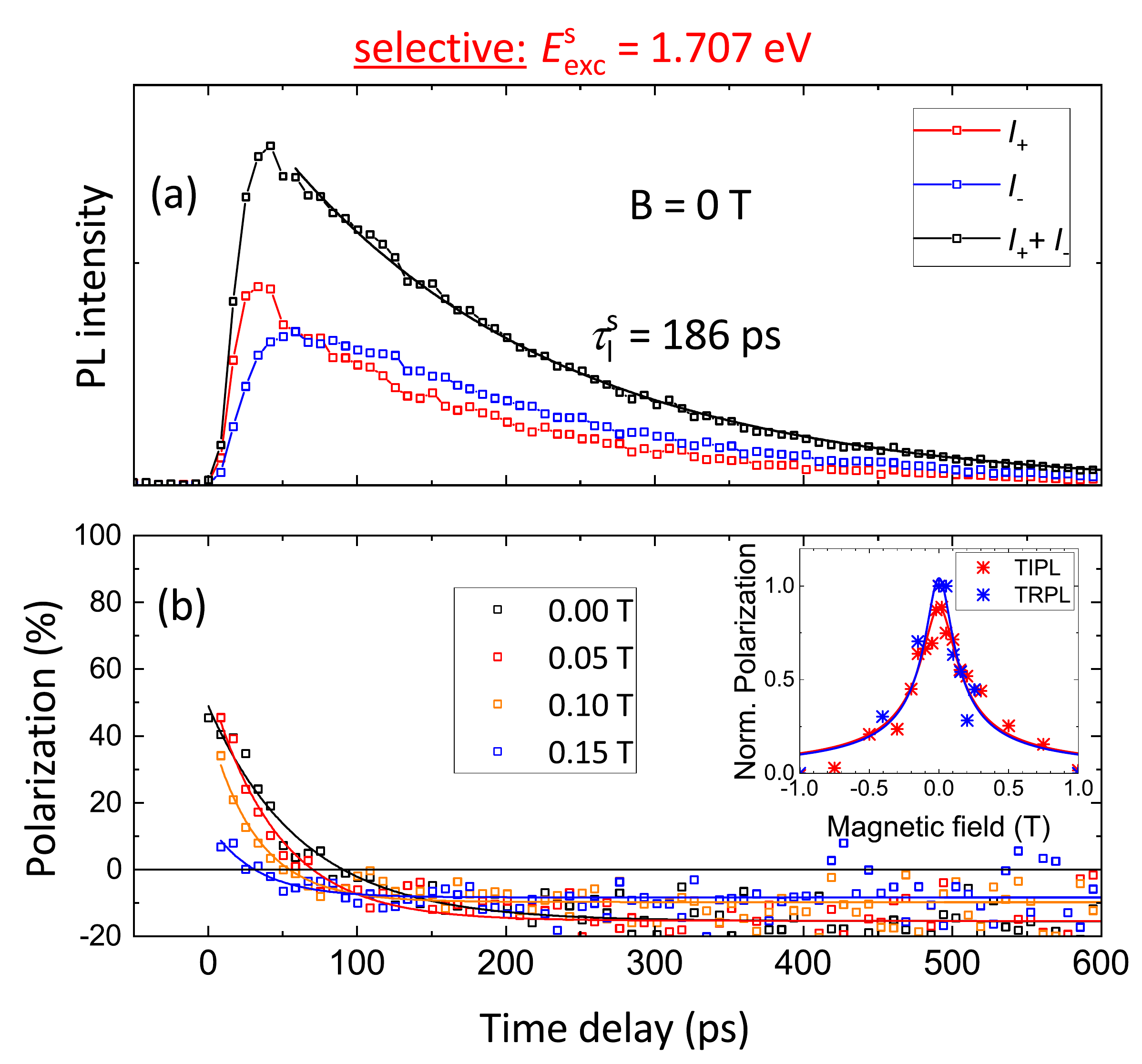}
\caption{\label{TRPL_RegimeII} (a) Time-resolved photoluminescence (TRPL) measurements using $E^\text{s}_\text{exc} = 1.707\,$eV excitation energy at $T=1.6\,$K. The evaluated carrier relaxation time corresponding to the exponential regression reads $\tau^\text{s}_\text{l} = 186 \pm 3$ ps. (b) Time-resolved polarization measurements. The colored lines are fits using Eq.~(\ref{eq:Time_dep_OO}). The parameters of all fits are listed in Table~\ref{tab:Negative_Regime}. The inset compares normalized depolarization curves for time-integrated photoluminescence (TIPL) and time-resolved photoluminescence (TRPL) measurements at various magnetic fields. Both sets of measurements possess similar FWHM of 0.4\,T, which points to their common origin.}
\end{figure}

Figure~\ref{TRPL_RegimeII} presents TR measurements of the PL and the OO in case of selective excitation. 
Similar to the analysis for nonselective excitation, Fig.~\ref{TRPL_RegimeII}(a) shows transient PL intensities for $\sigma^+$ and $\sigma^-$ polarized detection. The respective trion lifetime reads $\tau_\text{l}^\text{s} = 186\,$ps and is roughly twice longer than for nonselective excitation. 
The main feature in Fig.~\ref{TRPL_RegimeII}(a) is the crossing of the transients for $\sigma^+$ and $\sigma^-$ polarized excitation. 
The crossing of the red and blue curves occurs around 60\,ps and indicates a sign inversion of the OO. 

\begin{table}
\begin{tabular}{|c|c|c|c||c|}
\hline
$B$ (T) & $P^\text{s}_0$ (\%)   & $\tau^\text{s}_\text{s}$ (ps) & $P^\text{s}_\infty$ (\%) & $\tau^\text{s}_\text{l}$ (ps)\\ \hline
0.00    & 64 $\pm$ 4 & 63 $\pm$ 7        & $-$(15 $\pm$ 1) & 190 $\pm$ 4 \\ 
0.05  & 53 $\pm$ 4 & 65 $\pm$ 9        & $-$(16 $\pm$ 1) & 180 $\pm$ 3 \\ 
0.10   & 53 $\pm$ 9 & 32 $\pm$ 7        & $-$(10 $\pm$ 1)& 172 $\pm$ 3 \\
0.15  & 22 $\pm$ 8 & 32 $\pm$ 15        & $-$(8 $\pm$ 1) & 165 $\pm$ 2 \\ \hline
\end{tabular}
\caption{Evaluated parameters of the fits shown in Fig.~\ref{TRPL_RegimeII}.}
\label{tab:Negative_Regime}
\end{table}

Figure~\ref{TRPL_RegimeII}(b) presents time-resolved measurements of the OO at different magnetic fields of $B =$ 0, 0.05, 0.10 and $0.15\,$T. 
Similar to the observation for nonselective excitation, the OO amplitude possesses a large positive value of about 40\,\% at $B=0\,$T. 
The OO exponentially decays until it asymptotically saturates at a negative value; i.e. $P^\text{s}_\infty < 0$. The respective values are listed in Table~\ref{tab:Negative_Regime}. The constant negative value of $P^\text{s}_\infty$ suggests a contribution to the OO, which is not purely given by the dynamics of photogenerated carriers. Instead, an additional contribution possessing a very long spin relaxation time is expected. 

In TIPL measurements the observed OO is integrated over long times. 
Since the negative contribution to the OO in Fig.~\ref{TRPL_RegimeII}(b) is larger than the positive one, a negative value of the OO is expected in time-integrated measurements and, indeed, observed in Fig.~\ref{Polarization_Voigt}.
The inset in Fig.~\ref{TRPL_RegimeII}(b) shows normalized depolarization curves for TIPL and TRPL. 
Both measurements possess a similar FWHM of 0.4\,T, which links to their common origin.

\begin{figure}[t]
\includegraphics[width=0.48\textwidth]{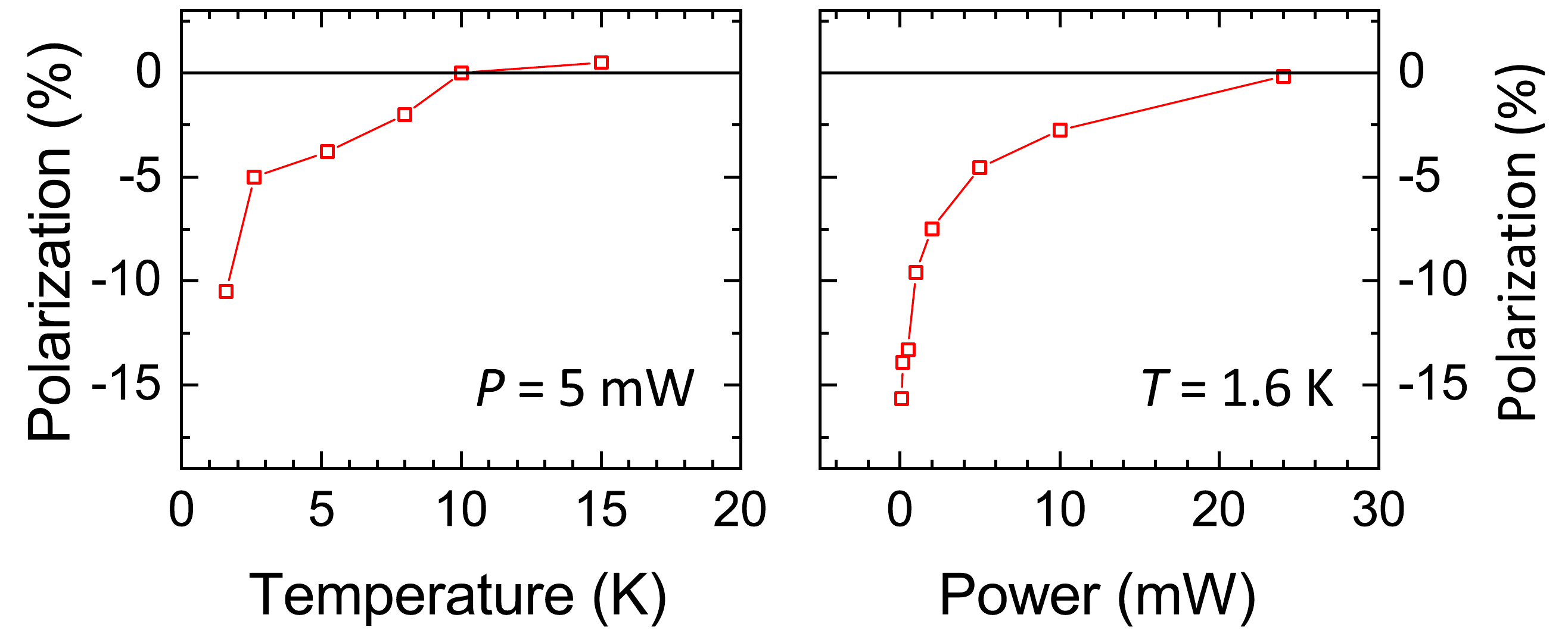}
\caption{\label{Temp_dep_OO} (a) Temperature and (b) power dependence of the time-integrated degree of OO at $B = 0\,$T for $E^\text{s}_\text{exc} = 1.707\,$eV. The detection energy reads $E_\text{det} = 1.705\,$eV. }
\end{figure}

The influence of the lattice temperature and excitation power is shown in Figs.~\ref{Temp_dep_OO}(a) and \ref{Temp_dep_OO}(b), respectively. 
It is clearly seen that an increasing lattice temperature suppresses the negative OO. 
A similar suppression of the negative OO is obtained by an increase of the excitation power, see Fig.~\ref{Temp_dep_OO}(b).

To summarize the experimental results at low temperatures, we observe a positive OO under nonselective excitation (close to the X state), while we find a negative OO under selective excitation of the T$^+$ state, see Fig.~\ref{Exc_energy_dep_OO}. 
By increasing the transverse magnetic field, we observe depolarization curves in both regimes, whose widths significantly differ, see Fig.~\ref{Polarization_Voigt}. The  depolarization curves may hint to the Hanle effect, but time-resolved measurements reveal that there are no oscillations present for non-zero magnetic fields, see Figs.~\ref{TRPL_RegimeI} and \ref{TRPL_RegimeII}. 
Note, that such oscillations are expected for the Hanle effect which is based on the Larmor precession of the mean spin in the transverse magnetic field (see e.g. \cite{meier_optical_1984}). Hence, the depolarization curves cannot originate from the Hanle effect. 

\begin{figure*}
\includegraphics[width=0.85\textwidth]{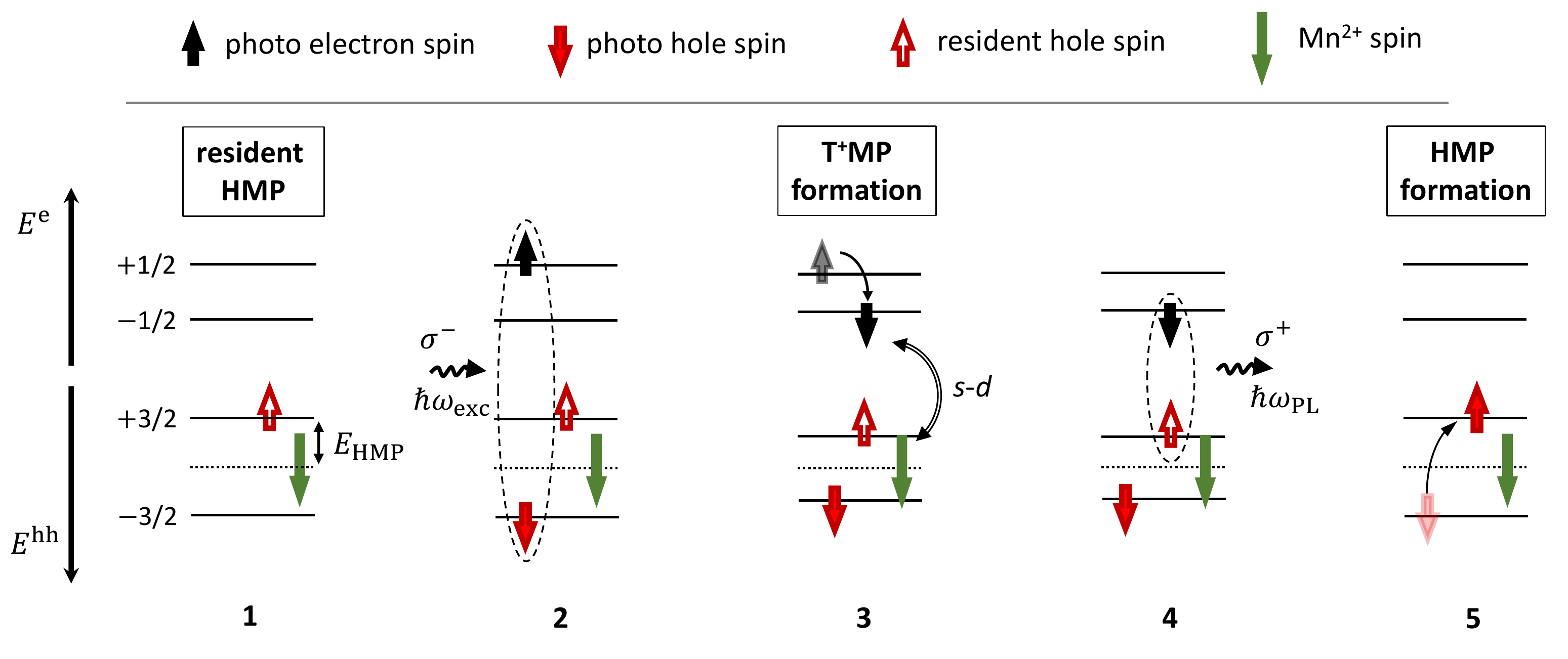}
\caption{\label{Trion_formation} Schematics of the trion magnetic polaron formation (namely T$^+$MP) from the hole magnetic polaron (HMP) under $\sigma^-$ polarized optical excitation. It explains the effect of negative optical orientation under selective excitation of trions (T$^+$). Note that the external magnetic field is assumed to be zero and all splittings of the spin states are provided by the exchange interactions of involved carriers with polarized Mn spins.}
\end{figure*}

Instead of oscillations, for selective excitation we observe a sign inversion of the OO, which happens at about 60\,ps delay time, see Fig.~\ref{TRPL_RegimeII}. 
The subsequent constant negative OO, however, lasts for very long times. The sign inversion of the OO hints to a spin-flip process of the photogenerated electron-hole pair, while the constant negative value indicates a very long spin relaxation time. 
Hence, we are looking for excitations, which are able to switch and hold the OO. Suitable candidates are hole magnetic polarons (HMPs), which are bound states of resident heavy holes surrounded by magnetic ions, i.e. Mn$^{2+}$. 
We propose that the interaction of photogenerated electron-hole pairs and resident hole magnetic polarons result in the formation of trion magnetic polarons (T$^+$MPs), which formation is accompanied by a sign inversion of the OO. The proposed mechanism was reported  by Zhukov et al.~\cite{zhukov_optical_2019} and results in the OO of resident HMPs.


\section{\label{sec:model_OO} Formation of trion magnetic polaron from hole magnetic polaron}

In Fig.~\ref{Trion_formation} we present schematically the process of trion magnetic polaron formation under selective excitation which photogenerates positively charged trions. Correspondingly, this kind of magnetic polaron will be further abbreviated T$^+$MP. In our experiments, the T$^+$MP evolves from the hole magnetic polaron (HMP) formed by a resident hole. The T$^+$MP formation results in reducing the Mn spin polarization and is accompanied by spin flips of the involved electrons. Here, we consider the case of zero external magnetic field, where all splittings of spin states are provided by the exchange interactions of the involved carriers with polarized Mn$^{2+}$ ions. The presented model allows us to explain the main experimental findings: (i) the inverted sign of optical orientation and (ii) the negative optical orientation, persisting for long times, see Fig.~\ref{TRPL_RegimeII}(b). 

The mechanism of the T$^+$MP formation has several stages shown in Fig.~\ref{Trion_formation}. It is more convenient to consider the case when $\sigma^-$ polarized light is used for excitation, while all conclusions are also valid for $\sigma^+$ excitation. 

Stage 1: Before photoexcitation the QW hosts resident holes, which form HMPs. The open red arrow represents the spin of the resident hole and the bold green arrow shows the Mn$^{2+}$ spin polarization within the HMP. The arrows are oriented opposite to each other due to the negative sign of the $p$-$d$ exchange interaction constant ($\beta < 0$)~\cite{gaj_introduction_2010}. Due to the anisotropy of the heavy-hole $g$-factor in narrow QWs ($g_{zz} \gg g_{xx},\, g_{yy}$) \cite{merkulov_two-dimensional_1995}, the hole spin is fixed along the structure growth axis ($z$-axis). Note, that at zero magnetic field the HMP orientations are equally distributed along and opposite to the $z$-axis, but the helicity of the exciting photons makes one of these directions favorable for photogeneration of a positively charged trion in its ground state, and, therefore, for the T$^+$MP formation. This favorable orientation is shown in Fig.~\ref{Trion_formation}. The HMP binding energy $E_\text{HMP}$ is the energy difference between the hole energy in an environment of unpolarized Mn spins (dashed line) and of polarized Mn spins in the hole exchange field.  

Stage 2: Resonant excitation with $\sigma^-$ circular polarization photogenerates electrons with spin $+1/2$ (black arrow) and holes with spin $-3/2$ (closed red arrow). If generation of such a pair happens in the vicinity of a resident hole with spin $+3/2$, this will result in generation of a positively charged trion T$^+$ ($+1/2,-3/2,+3/2$), where the two holes have opposite spins ($-3/2,+3/2$). The photon energy for this initial photoexcitation is $\hbar \omega_\text{exc}$. It is important that this trion is exposed to the exchange field of Mn spins, that was formed within the HMP. This location provides the spin splitting of the hole and electron states. The electron with spin $+1/2$ is not the lowest in energy state, which has spin $-1/2$.

Stage 3: Here, the trion magnetic polaron is formed. We base our qualitative description here on the rigorous theoretical analysis given below and the results of model calculations for the case of (Cd,Mn)Te QWs with low Mn concentration. Apparently, if the two holes in T$^+$ were in the singlet state, which is the ground trion state, the mean values of their spins would be zero, as would be the exchange fields that the holes exert on the Mn spins. But actually, since  the two holes have different localization volumes, the admixture of singlet and triplet holes states provides an effective hole exchange field. The polarization of the Mn spins by this exchange field creates the Mn exchange field that maintains the singlet-triplet mixing. The hole exchange field within T$^+$MP is smaller than that within the HMP, which is represented in Fig.~\ref{Trion_formation} by decreasing spin splittings at stage 3 compared to stage 1. In this way, the trion magnetic polaron is formed. It is worthwhile to note, that this is a rather unusual physical situation for magnetic polarons, where the polaron formation process is accompanied by a decrease of the Mn polarization and not by an increase. Also, during the T$^+$MP formation the electron experiences a spin-flip and ends up in its lowest spin state $-1/2$ within the vicinity of the polarized Mn spins. 

Stage 4:  The electron $-1/2$ recombines with the resident hole $+3/2$, leading to emission of a $\sigma^+$ circularly polarized photon with energy $\hbar \omega_\text{PL}$. Note, that the emission polarization is opposite to the excitation one, as we observed in experiment. The photoexcited hole $-3/2$ is left after the trion recombination and becomes a resident hole. The Stokes shift between the excitation and emission photon energies $\hbar \omega_\text{exc}-\hbar \omega_\text{PL}$ provides information about the exchange energies of electron and hole in both T$^+$MP and HMP. In the case when the T$^+$MP formation time exceeds its recombination time, the trion polaron can recombine before being fully developed, which needs to be taken into account when evaluating the exchange energies from experimental data.  

Stage 5: The hole, which is left after the trion recombination with spin $-3/2$, flips its spin to $+3/2$ to be in an energetically favorable orientation with the polarized Mn spins. Then it increases the Mn polarization back to the extent that it has in the HMP. The process ends up with a HMP formed and with the same hole spin orientation as at the start in stage 1.

\section{\label{sec:theory} Theory of trion magnetic polaron}

Charge carriers in the valence and conduction band and the $d$-electrons of the Mn ions in diluted magnetic semiconductors are coupled by exchange interaction. In case of localized carriers, the interaction energy due to this interaction can be expressed as:
\begin{align}
E^\text{e}_\text{ex} = -\alpha N_0 x s_z \int S_z(\boldsymbol{r}) \Psi^2_\text{e} (\boldsymbol{r}) \text{d}^3 r, 
\label{eq:E^e}
\end{align}
for the case of an electron, where $\alpha N_0$ is the $s$-$d$ exchange constant (in Cd$_{1-x}$Mn$_x$Te $\alpha N_0 = 220$~meV) and $x$ is the Mn$^{2+}$ concentration, $N_0$ is the concentration of cation sites in the crystal. Further, $S_z (\boldsymbol{r})$ is the $z$-projection of the mean spin of the magnetic ions, $\Psi_\text{e}(\boldsymbol{r})$ is the envelope wave function of the localized electron. A similar equation is obtained for the exchange energy of a hole and the system of magnetic ions:
\begin{align}
E^\text{h}_\text{ex} = -\frac{1}{3} |\beta| N_0 x J_z \int S_z(\boldsymbol{r}) \Psi^2_\text{h} (\boldsymbol{r}) \text{d}^3 r \,,
\end{align} 
where $\beta N_0 =-880$~meV is the $p$-$d$ exchange constant and $\Psi_\text{h}(\boldsymbol{r})$ is the envelope function of the hole.

If localized resident holes are present in the QW, they form hole magnetic polarons with magnetic moments oriented along the structure growth axis $z$. Under optical excitation, an exciton created by light absorption can be captured by or directly excited to the potential trap which is already occupied by a resident hole. As a result, a localized trion is formed. In the following, we consider the magnetic polaron formed by such a trion, namely T$^+$MP.

The knowledge accumulated to date on exciton magnetic polarons in similar structures \cite{yakovlev_magnetic_2010} allows us to make several important assumptions, that greatly simplify the theoretical consideration of T$^+$MP. Firstly, as in (Cd,Mn)Te the energy constant of the $s$-$d$ exchange interaction is four times smaller than that of the $p$-$d$ interaction, we neglect the polarization of the Mn spins due to their interaction with the electrons. With this assumption, we can consider the magnetic polaron formed by the two holes and then calculate its effect on the electron spin polarization. Secondly, we neglect the admixture of light-hole states to the localized heavy holes that form the T$^+$MP, and therefore assume that the $p$-$d$ exchange interaction involves only the $z$-projections of hole spins. Thirdly, we describe the Mn spin polarization within the linear response approximation. 

We further assume that, in contrast to quantum dots, the Coulomb repulsion of the two holes is stronger than the localization potential of the trap. In this case, the second hole cannot occupy the same orbital state as the resident one, and will have a larger localization radius. Since the two holes have the same Bloch amplitudes of the heavy-hole subband, the energy spectrum of their exchange Hamiltonian consists of a singlet and a triplet~\cite{kavokin_symmetry_2004}. The hole-hole exchange interaction can therefore be represented by a scalar product of pseudospin $1/2$ vectors $j$ defined such that $j_z=J_z/3$. The hole-hole exchange constant $\Delta_\text{hh}$ depends on the shape and depth of the trap. We assume it to be proportional to the squared overlap integral $I$ of the hole wave functions:
\begin{align}
\Delta_\text{hh}=\Delta_0I^2 \,,
\label{eq:app:Hole-hole exchange}
\end{align}
where $\Delta_0$ is a characteristic energy of the order of several meV.

On the opposite, the $p$-$d$ exchange interaction of heavy holes with manganese is anisotropic, so that it affects only the $z$-projection of $j$. The total exchange Hamiltonian of T$^+$MP can, therefore, be written as: 
\begin{align}
\hat{H}_\text{p} = 2 \Delta_\text{hh} \boldsymbol{j}_1 \boldsymbol{j}_2 + E_\text{p1} j_{1z} + E_\text{p2} j_{2z} \,.
\label{eq:app:Hamiltonian_polaron}
\end{align}
Here
\begin{align}
E_\text{p1} = \beta N_0 x \int S_z \Psi^2_\text{h1} (\boldsymbol{r}) \text{d}^3 r,  \nonumber \\
E_\text{p2} = \beta N_0 x \int S_z \Psi^2_\text{h2} (\boldsymbol{r}) \text{d}^3 r,  \label{eq:Ep}
\end{align}
where $\Psi_\text{h1}$ and $\Psi_\text{h2}$ are the envelope wave functions of the two localized hole states.

In the case when the Mn spins are not polarized (i.e., $S_z=0$ and therefore $E_\text{p1} = E_\text{p2} = 0$), the eigenstates of the exchange Hamiltonian correspond to the values of the total pseudospin $\boldsymbol{F} = \boldsymbol{j}_1 + \boldsymbol{j}_2$: $F = 0$ (singlet, $\frac{1}{\sqrt{2}} \left( \uparrow \downarrow - \uparrow \downarrow \right)$) and $F = 1$ (triplet) with $z$-projections $F_z = \pm 1 \left( \uparrow \uparrow \text{and} \downarrow \downarrow \right)$ and $F_z = 0 \left(  \frac{1}{\sqrt{2}} \left( \uparrow \downarrow + \uparrow \downarrow \right)  \right)$. If $E_\text{p1}$ and/or $E_\text{p2}$ are not zero, the matrix of the Hamiltonian in Eq.~(\ref{eq:app:Hamiltonian_polaron}) using the just introduced basis states of $F$ is shown in Tab.~\ref{tab:app:Matrix_F_states}.

\begin{table}
\begin{center}
\begin{tabular}{ l| c c c c }
						&	$F = 1,+1$ 		& $F = 1,0$			& $F = 1,-1$		& $F = 0$\\ \hline
$F = 1,+1$ 		& $\frac{\Delta_\text{hh}}{2} + \frac{E_\text{p1} + E_\text{p2}}{2}$ 							& 0 							& 0 							& 0 \\
$F = 1,0$ 		& 0						& $\frac{\Delta_\text{hh}}{2}$ 	 							& 0 							& $\frac{E_\text{p1} - E_\text{p2}}{2}$ \\
$F = 1,-1$		&0						& 0 							& $\frac{\Delta_\text{hh}}{2} - \frac{E_\text{p1} + E_\text{p2}}{2}$ 	 							& 0 \\ 
$F = 0$			&0						& $\frac{E_\text{p1} - E_\text{p2}}{2} $ 							& 0 	 							& $-\frac{3\Delta_\text{hh}}{2}$   
\end{tabular}
\end{center}
\caption{Matrix of the Hamiltonian in Eq.~(\ref{eq:app:Hamiltonian_polaron}) using the basis states of $F$.\label{tab:app:Matrix_F_states}}
\end{table}
Diagonalization of the matrix in Tab.~\ref{tab:app:Matrix_F_states} yields two triplet states, which are not mixed, $\ket{F = 1, +1}$ and $\ket{F = 1, -1}$ with energies $E_{1+} = \frac{\Delta_\text{hh}}{2} + \frac{E_\text{p1} + E_\text{p2}}{2}$ and $E_{1-} = \frac{\Delta_\text{hh}}{2} - \frac{E_\text{p1} + E_\text{p2}}{2}$, and two mixed states resulting from mixing of the singlet and triplet state with zero $z$-projection of $F$:
\begin{align}
\ket{+0} &= \ket{F = 1, 0} \cos \frac{\varphi}{2} + \ket{F = 0} \sin \frac{\varphi}{2}, \\
\ket{-0} &= \ket{F = 0} \cos \frac{\varphi}{2} - \ket{F=1,0} \sin \frac{\varphi}{2} ,
\end{align}
where the angle $\varphi$ is defined by the relations
\begin{align}
\cos \varphi &= \frac{2 \Delta_\text{hh}}{\sqrt{4 \Delta^2_\text{hh} + \left( E_\text{p1}  - E_\text{p2}  \right)^2}}, 
\end{align}
\begin{align}
\sin \varphi &= \frac{E_\text{p1}  - E_\text{p2}}{\sqrt{4 \Delta^2_\text{hh} + \left( E_\text{p1}  - E_\text{p2}  \right)^2}}.
\label{eq:sin_varphi}
\end{align}
The corresponding energies are
\begin{align}
E_{+0} = -  \frac{\Delta_\text{hh}}{2} +\frac{1}{2} \sqrt{4 \Delta^2_\text{hh} + \left( E_\text{p1}  - E_\text{p2}  \right)^2}, \\
E_{-0} = -  \frac{\Delta_\text{hh}}{2} - \frac{1}{2}\sqrt{4 \Delta^2_\text{hh} + \left( E_\text{p1}  - E_\text{p2}  \right)^2}. \label{eq:E-0}
\end{align}
The mean spin of Mn, induced by the interaction with the hole spins, in the linear-response approximation equals to
\begin{align}
S_z(\boldsymbol{r}) = \beta \frac{S(S+1)}{3 k_\text{B} T^*} \left[ \braket{j_{1z}} \Psi^2_\text{h1}(\boldsymbol{r}) + \braket{j_{2z}} \Psi^2_\text{h2}(\boldsymbol{r}) \right],
\label{eq:app:Mn_spin}
\end{align}
where $T^*$ is the Mn effective temperature, $k_\text{B}$ is the Boltzmann constant, and $S=5/2$ is the Mn spin. 

Here, for the ground state
\begin{align}
\braket{j_{1z}} &= \braket{-0|j_{1z}|-0} = \cos \frac{\varphi}{2} \sin \frac{\varphi}{2} = \frac{1}{2} \sin \varphi, \nonumber \\
\braket{j_{2z}} &= \braket{-0|j_{2z}|-0} = -\cos \frac{\varphi}{2} \sin \frac{\varphi}{2} = -\frac{1}{2} \sin \varphi. \label{eq:j}
\end{align}
From Eqs.~(\ref{eq:app:Mn_spin}) and (\ref{eq:j}), we find
\begin{align}
S_z(\boldsymbol{r}) = \beta \frac{S(S+1)}{3 k_\text{B} T^*} \left[\Psi^2_\text{h1}(\boldsymbol{r}) -\Psi^2_\text{h2}(\boldsymbol{r}) \right] \frac{1}{2} \sin \varphi \,.
\end{align}
Then, from Eq.~(\ref{eq:Ep}) it follows that
\begin{align}
E_\text{p1}  - E_\text{p2} 
	&= \beta^2 N_0 x \frac{S(S+1)}{6 k_\text{B} T^*} \sin \varphi \, \times \nonumber\\
	 & \qquad \int \left[\Psi^2_\text{h1}(\boldsymbol{r}) -\Psi^2_\text{h2}(\boldsymbol{r}) \right]^2\text{d}^3 r .
	 \end{align}
By substituting $\sin \varphi$ from Eq.~(\ref{eq:sin_varphi}) we get	 
	\begin{align}	 
E_\text{p1}  - E_\text{p2} &= \beta^2 N_0 x \frac{S(S+1)}{6 k_\text{B} T^*} \frac{E_\text{p1}  - E_\text{p2}}{\sqrt{4 \Delta^2_\text{hh} + \left( E_\text{p1}  - E_\text{p2}  \right)^2}} \times \nonumber \\ 
& \qquad \int \left[\Psi^2_\text{h1}(\boldsymbol{r}) -\Psi^2_\text{h2}(\boldsymbol{r}) \right]^2\text{d}^3 r .
 \label{eq:app:B_p1_minus_B_p1}
\end{align}
Solving Eq.~(\ref{eq:app:B_p1_minus_B_p1}) with respect to $E_\text{p1}  - E_\text{p2}$, one can find the conditions for T$^+$MP formation, and ultimately its energy and angular momentum. The non-trivial (polaron) solution is
\begin{align}
\left( E_\text{p1}  - E_\text{p2} \right)^2 = \nonumber \\
	= \left( \beta^2 N_0 x \frac{S(S+1)}{6 k_\text{B} T^*} \int \left[\Psi^2_\text{h1}(\boldsymbol{r}) -\Psi^2_\text{h2}(\boldsymbol{r}) \right]^2\text{d}^3 r  \right)^2 
	  - 4 \Delta^2_\text{hh}. 
\label{eq:app:B_p1_minus_B_p1_q}
\end{align}
The T$^+$MP stability condition reads 
\begin{align}
\beta^2 N_0 x \frac{S(S+1)}{12 k_\text{B} T^*\Delta_\text{hh}} \int \left[\Psi^2_\text{h1}(\boldsymbol{r}) -\Psi^2_\text{h2}(\boldsymbol{r}) \right]^2\text{d}^3 r  > 1 \,.
\end{align}
If this condition is satisfied, it follows from Eqs.~(\ref{eq:E-0}) and (\ref{eq:app:B_p1_minus_B_p1_q}) that the ground-state energy of the two holes interacting with each other and with the Mn spins equals
\begin{align}
E_{-0}^{\text{TMP}}= - \frac{\Delta_{\rm hh}}{2} - \beta^2 N_0 x \frac{S(S+1)}{12 k_\text{B} T^*} \int \left[\Psi^2_\text{h1}(\boldsymbol{r}) -\Psi^2_\text{h2}(\boldsymbol{r}) \right]^2\text{d}^3 r.\label{eq:E_TMP_0}
\end{align}
If this condition is not satisfied, the ground-state energy is
\begin{align}
E_{-0}^{\text{no TMP}}= - \frac{3\Delta_{\rm hh}}{2} \,.
\end{align}

The absolute value of the difference between these two energies constitutes the binding energy of T$^+$MP (here we neglect the electron contribution):
 \begin{align}
E_{\text{TMP}}= \beta^2 N_0 x \frac{S(S+1)}{12 k_\text{B} T^*} \int \left[\Psi^2_\text{h1}(\boldsymbol{r}) -\Psi^2_\text{h2}(\boldsymbol{r}) \right]^2\text{d}^3 r-\Delta_{\rm hh}.\label{eq:E_TMP}
\end{align}

Finally, we get the Mn mean spin distribution within the T$^+$MP: 
\begin{align}
S_z(\boldsymbol{r})=\frac{ \beta S(S+1)}{6 k_\text{B} T^*} \sin \varphi 
	 \left[\Psi^2_\text{h1}(\boldsymbol{r}) -\Psi^2_\text{h2}(\boldsymbol{r}) \right],  
\label{eq:S_z} 	
\end{align}
where, according to Eqs.~(\ref{eq:sin_varphi}), (\ref{eq:app:B_p1_minus_B_p1_q}), and (\ref{eq:E_TMP})
\begin{align}
 \sin \varphi =\sqrt{1-\left( \frac{\Delta_{\rm hh}}{\Delta_{\rm hh}+E_{\text{TMP}}}  \right)^2} \,.
\label{eq:sin}
\end{align}

Combining Eqs.~(\ref{eq:E^e}) and (\ref{eq:S_z}), we obtain the expression for the electron energy in the T$^+$MP:
\begin{align}
E^\text{e}_\text{ex} = \frac{|\alpha \beta| N_0 xS(S+1)}{12 k_\text{B} T^*} \sin \varphi 
 \int \Psi^2_\text{e}(\boldsymbol{r}) |\Psi^2_\text{h1}(\boldsymbol{r}) -\Psi^2_\text{h2}(\boldsymbol{r}) | \text{d}^3 r \,.
  \label{eq:E^e final}
\end{align}

When T$^+$MP is formed under resonant optical excitation, the experimentally measured characteristic quantity of its energy is the Stokes shift between the photon energies of excitation and emission. 
\begin{align}
 \Delta E_\text{S} = \hbar \omega_{\rm exc} - \hbar \omega_{\rm PL} \,. 
\end{align}
From energy conservation
\begin{align}
 \hbar \omega_{\text{exc}}+E_{i} = \hbar \omega_{\text{PL}} + E_{f} + \Delta E_{\text{phonon}},
\end{align}
where $\Delta E_i$ is the energy of the Mn spins and the resident hole before excitation, $E_f$ is the energy of the Mn spins and the resident hole right after emission of the PL photon, and $\Delta E_\text{phonon}$ is the amount of energy transferred to (or from) phonons during T$^+$MP formation. Therefore, 
\begin{align}
\Delta E_\text{S} = E_f -E_i +\Delta E_\text{phonon}.
\end{align}

The theory of T$^+$MP developed above allows us to determine all these energies. To this end, we replace the indices h1 and h2 of the hole wave functions with rh and ph, which denote resident hole and photoexcited hole, respectively.

The initial energy $E_i$ is, obviously, the binding energy of the HMP formed by a resident hole, while $E_f$ is the energy of the resident hole with the spin antiparallel (empty arrow in Fig.~\ref{Trion_formation}, stage 5) to the exchange field of the Mn ions left after T$^+$MP recombination: 
\begin{align}
\label{eq:E_HMP}
E_i  = E_{\text{HMP}}= - \beta^2 N_0 x \frac{S(S+1)}{12 k_\text{B} T^*}  \int \Psi^4_\text{rh}\text{d}^3 r .
	 \end{align}
\begin{align}
E_f=+ \beta^2 N_0 x \frac{S(S+1)}{12 k_\text{B} T^*} \sin \varphi \int \Psi^2_\text{rh}\left[\Psi^2_\text{rh}(\boldsymbol{r}) -\Psi^2_\text{ph}(\boldsymbol{r}) \right]\text{d}^3 r .
	 \end{align}

The energy transferred to/from phonons equals to the difference of the ground state energies of trion and magnetic ions right after its optical excitation and after formation of the T$^+$MP: 
\begin{equation}
\Delta E_{\text{phonon}}=E^{\rm T}_{-0}-E^{\rm TMP}_{-0} ,  
\end{equation}
where 
\begin{align}
E^\text{T}_{-0} = -  \frac{\Delta_\text{hh}}{2} - \frac{1}{2}\sqrt{4 \Delta^2_\text{hh} + \left( E_\text{p1}  - E_\text{p2}  \right)^2},
\end{align}
and 
\begin{align}
 E_\text{p1}  - E_\text{p2}=\beta^2 N_0 x \frac{S(S+1)}{6 k_\text{B} T^*}  
	  \int \Psi^2_\text{rh}\left[\Psi^2_\text{rh}(\boldsymbol{r}) -\Psi^2_\text{ph}(\boldsymbol{r}) \right]\text{d}^3 r .
	 \end{align}
These expressions are obtained using Eq.~(\ref{eq:E-0}) and Eq.~(\ref{eq:Ep}) with $S_z(\boldsymbol{r})$ defined by the exchange field of the resident hole.

The energy of the formed T$^+$MP is given by Eq.~(\ref{eq:E_TMP_0}):
\begin{align}
E_{-0}^{\text{TMP}}= - \frac{\Delta_{\rm hh}}{2} - \beta^2 N_0 x \frac{S(S+1)}{12 k_\text{B} T^*} \int \left[\Psi^2_\text{rh}(\boldsymbol{r}) -\Psi^2_\text{ph}(\boldsymbol{r}) \right]^2\text{d}^3 r.
\end{align}

Finally, we get: 
\begin{align}
\Delta E_{\text{S}} =  \beta^2 N_0 x \frac{S(S+1)}{12 k_\text{B} T^*}  \times \nonumber \\
\left[ \sin \varphi \int\Psi^2_\text{rh}(\boldsymbol{r}) \left[\Psi^2_\text{rh}(\boldsymbol{r}) -\Psi^2_\text{ph}(\boldsymbol{r}) \right]\text{d}^3 r+ \int\Psi^4_\text{rh}\text{d}^3 r\right]- \nonumber \\
-\frac{1}{2}\sqrt{4\Delta^2_\text{hh}+\left[ \beta^2 N_0 x \frac{S(S+1)}{6 k_\text{B} T^*} \int \Psi^2_\text{rh}\left[\Psi^2_\text{rh} -\Psi^2_\text{ph} \right]\text{d}^3 r \right]^2} \nonumber \\
+\beta^2 N_0 x \frac{S(S+1)}{12 k_\text{B} T^*} \int \left[\Psi^2_\text{rh}(\boldsymbol{r}) -\Psi^2_\text{ph}(\boldsymbol{r}) \right]^2\text{d}^3 r .
\end{align}
Comparing with Eq.~(\ref{eq:E_TMP}), one can see that the Stokes shift does not coincide with the T$^+$MP binding energy, in contrast to an exciton magnetic polaron. 
In the next section, the energy parameters of the T$^+$MP will be calculated using model envelope functions of the electron and the two holes. 

\section{\label{sec:mod} Calculations with model envelope functions}

Since the exact form of the in-plane localization potential is not known, we use model envelope wave functions:
\begin{align}
\Psi_i = \sqrt{\frac{4}{\pi L_z a_i^2}} \cos \left( \frac{\pi}{L_z} z \right) \exp \left( - \frac{r}{a_i} \right) \,, 
\label{eq:functions}
\end{align}
where $r = \sqrt{x ^2 + y^2}$, $L_z$ is the QW width, and the $a_i$ are the respective radii for the electron or the holes ($i$ = e, rh, ph). 
Here, $z = 0$ corresponds to the center of the QW layer. 

Furthermore, we normalize the used radii with respect to the localization radii of the resident hole, i.e. introduce the parameters $\kappa_\text{e}  = a_\text{e} / a_\text{rh}$ and $\kappa = a_\text{ph} / a_\text{rh}$. That way, we get the following expressions for the T$^+$MP binding energy:
\begin{align}
E_{\text{TMP}}=\frac{S(S+1)\beta^2N_0x}{16\pi L_z a^2_{\rm rh}k_{\text{B}}T^*}\left[ 1+\frac{1}{\kappa^2}-\frac{8}{(1+\kappa)^2}\right] \nonumber \\
-16\Delta_0\frac{\kappa^2}{(1+\kappa)^4}, \label{eq:E_TMPkappa}
\end{align}
as well as for the hole-hole exchange constant:
\begin{align}
\Delta_{\rm hh}=16\Delta_0\frac{\kappa^2}{\left(1+\kappa \right)^4}. \label{eq:Delta_hh}
\end{align}
The electron exchange energy in the T$^+$MP is:
\begin{align}
 E^\text{e}_\text{ex} = \frac{|\alpha \beta| N_0 xS(S+1)}{4 \pi L_z a^2_{\rm rh} k_\text{B} T^*} \sin \varphi \left[\frac{1}{(1+\kappa_\text{e})^2}-\frac{1}{(\kappa+\kappa_\text{e})^2} \right]
\end{align}
and the Stokes shift of the T$^+$MP is:
\begin{align}
\Delta E_{\text{S}} =  \beta^2 N_0 x \frac{S(S+1)}{16 \pi L_z a^2_{\rm rh} k_\text{B} T^*} \left[\sin \varphi   
\left(1-\frac{4}{(1+\kappa)^2}\right)+1  \right] - \nonumber \\
\frac{1}{2}\sqrt{4\Delta^2_\text{hh}+\left[ \beta^2 N_0 x \frac{S(S+1)}{8 \pi L_z a^2_{\rm rh} k_\text{B} T^*}  \left(1-\frac{4}{(1+\kappa)^2} \right) \right]^2} + \nonumber \\
\beta^2 N_0 x \frac{S(S+1)}{16 \pi L_z a^2_{\rm rh} k_\text{B} T^*} \left(1-\frac{8}{(1+\kappa)^2} + \frac{1}{\kappa^2} \right).
\label{eq:Stokes_kappa}
\end{align}
Here $\sin \varphi$ is obtained by inserting $E_{\text{TMP}}$ from Eq.~(\ref{eq:E_TMPkappa}) and $\Delta_{\rm hh}$ from Eq.~(\ref{eq:Delta_hh}) into Eq.~(\ref{eq:sin}).

The binding energy of the hole magnetic polaron, which is a major part of $\Delta E_\text{S}$, is independent of $\kappa$ and $\kappa_e$ and equals to
\begin{align}
E_{\text{HMP}}=\frac{S(S+1)\beta^2N_0x}{16\pi L_z a^2_{\rm rh}k_{\text{B}}T^*}. \label{eq:E_HMPkappa}
\end{align}

\section{\label{sec:model_Depol} Model of depolarization in a transverse  Magnetic Field}

When an external magnetic field $\boldsymbol{B}$ is applied parallel to the QW plane, it induces an additional polarization of the Mn spins: 
\begin{align}
S_{\rm Mn}(B) 
	&=\frac{ S(S+1)}{3 k_\text{B} T^*} \mu_{\rm B} g_\text{Mn} B.	
\end{align}
This field-induced mean Mn spin is perpendicular to the hole-induced mean Mn spin $S_z$ given by Eq.~(\ref{eq:S_z}).
The total energy of the electron spin interacting with the Mn spin system now reads  
\begin{align}
E^\text{e}_\text{ex} = \frac{|\alpha| N_0 x }{2} \sqrt{\left[\int \Psi^2_\text{e}(\boldsymbol{r}) S_z(\boldsymbol{r})d^3r \right]^2+S_{\rm Mn}(B)^2}= \nonumber \\
=\frac{|\alpha| N_0 x S(S+1)\mu_{\rm B} g_\text{Mn}}{6  k_\text{B} T^*} \sqrt{B_{\rm ex}^2+B^2} \,,
\label{eq:E^e field}
\end{align}
where
\begin{align} 
B_{\rm ex}=\frac{3  k_\text{B} T^*}{S(S+1)\mu_{\rm B} g_\text{Mn}}\int \Psi^2_\text{e}(\boldsymbol{r}) S_z(\boldsymbol{r})d^3r 
\label{eq:B_{ex}}
\end{align}
is the effective magnetic field of holes acting on the Mn spins in the T$^+$MP. 

At low temperature, the electon spin is fully polarized along the total effective field  $\boldsymbol{B}_\text{eff} = (B, 0, B_\text{ex})^T$. The angle between the effective field and the structure axis equals to $\theta=\arctan \frac{B}{B_\text{ex}}$. 

\begin{figure}[b]
\includegraphics[width=0.46\textwidth]{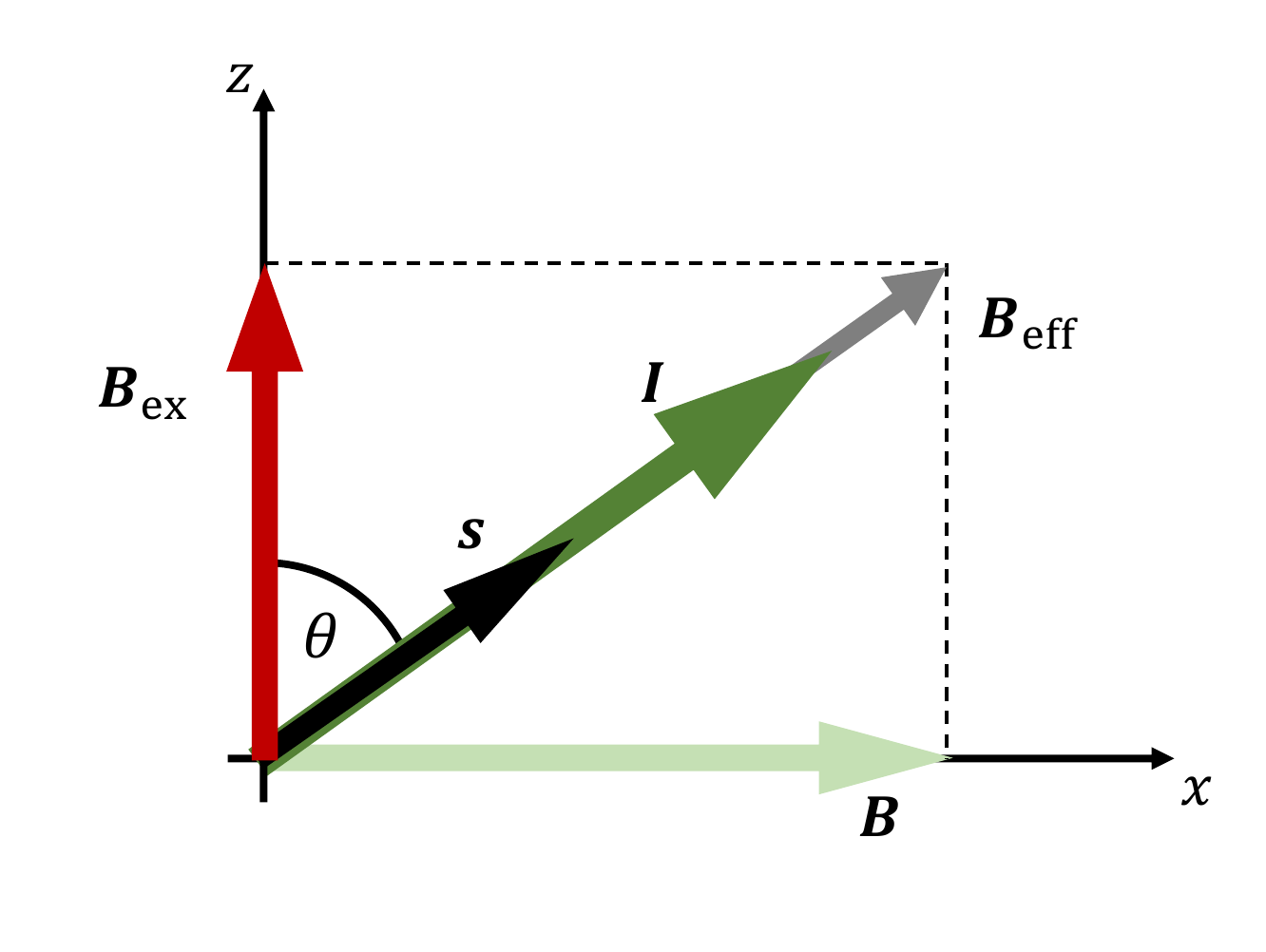}
\caption{\label{Mechanism_scheme} Scheme of the spin structure of the EMP in an external magnetic field applied in Voigt geometry. The electron spin is directed along the effective magnetic field $B_\text{eff} = \sqrt{B_\text{ex}^2 + B^2}$.}
\end{figure}

According to Fig.~\ref{Mechanism_scheme}, the Bloch vector describing the spin state of the electron can be written as
\begin{align}
\ket{\Psi} = \cos \left( \frac{\theta}{2} \right) \ket{+} + \sin \left( \frac{\theta}{2} \right) \ket{-},
\label{eq:Bloch_vector}
\end{align}
where $\ket{\pm}$ describe the basis vectors of the spin up and spin down state along the exchange field $B_\text{ex}$, respectively. 

For $\sigma ^-$  excitation, the intensities of the optical transitions with $\sigma ^-$  and $\sigma ^+$  polarization are proportional to the squared matrix elements for recombination of the electron with the photoexcited and the resident hole, respectively. Using Eq.~(\ref{eq:Bloch_vector}), we find for the PL circular polarization under optical orientation: 
 \begin{align}
P_{\rm oo} (B)
	&\propto  -\frac{I^2_{\rm e-rh}\cos^2\frac{\theta}{2}-I^2_{\rm e-ph}\sin^2\frac{\theta}{2}}{I^2_{\rm e-rh}\cos^2\frac{\theta}{2}+I^2_{\rm e-ph}\sin^2\frac{\theta}{2}}\propto \nonumber \\
\propto -\frac{\rho_{\rm eh}+\cos\theta}{1+\rho_{\rm eh}\cos\theta} \,,
	\label{eq:Pol_Bx}
\end{align}
where $\cos\theta=\frac{B_{\rm ex}}{\sqrt{B^2_{\rm ex}+B^2}}$ and $\rho_{\rm eh}=\frac{I^2_{\rm e-rh}-I^2_{\rm e-ph}}{I^2_{\rm e-rh}+I^2_{\rm e-ph}}$, while $I_{\rm e-rh}=\int \Psi_\text{e}(\boldsymbol{r})\Psi_\text{rh}(\boldsymbol{r})d^3r$ and $I_{\rm e-ph}=\int \Psi_\text{e}(\boldsymbol{r})\Psi_\text{ph}(\boldsymbol{r})d^3r$ are the overlap integrals of the electron with the resident and the photoexcited hole, respectively. Based on the model wave functions, used in the previous section, $I_{\rm e-rh}=\frac{4\kappa_{\rm e}}{(1+\kappa_{\rm e})^2}$ and $I_{\rm e-ph}=\frac{4\kappa_{\rm e}/\kappa}{(1+\kappa_{\rm e}/\kappa)^2}$.
In strong magnetic fields $B>>B_{\rm ex}$, $P_{\rm oo}(B)$ asymptotically reaches a constant level, that is proportional to $\rho_{\rm eh}$, which is determined by the difference of the overlap integrals of the electron with the two holes. In our experiments, this level is close to zero within the experimental accuracy, suggesting that the two overlap integrals are nearly equal to each other. Using the model wave functions, this condition is satisfied when $\kappa_{\rm e}=\sqrt{\kappa}$. Under this condition, the expression for the field dependence of the polarization simplifies to 
\begin{align}
P_{\rm oo} (B) \propto -\cos\theta \propto - \frac{B_{\rm ex}}{\sqrt{B^2_{\rm ex}+B^2}}.
	\label{eq:Pol_Bx_0}
\end{align} 
One can see from Fig.~\ref{Polarization_Voigt} that Eq.~(\ref{eq:Pol_Bx_0}) indeed describes the experimental dependence quite well.  
The depolarization curve can be used to extract the hole exchange field $B_\text{ex}$ from the curve's full width at half maximum (FWHM), $\Delta B$, according to
\begin{align}
B_\text{ex} = \frac{\Delta B}{2\sqrt{3}}\,.
\label{eq:FWHM_Bex}
\end{align}
A broader depolarization curve corresponds to a larger hole exchange field. Fit of the experimental data recorded for selective excitation in Fig.~\ref{Polarization_Voigt} gives $\Delta B=0.4$~T, which corresponds to $B_\text{ex}=0.12$~T.

In the following, we are interested in the exchange energy of the electron due to its interaction with the Mn spin polarization, which in turn is induced by the hole exchange field. In the absence of an external magnetic field the electron exchange energy can be approximated by \cite{zhukov_optical_2016}
\begin{align}
E_\text{ex}^\text{e}(B = 0) = \gamma^\text{e} B_\text{ex}\,,
\label{eq:electron_ex_energy}
\end{align}
where $\gamma^\text{e}$ = $\frac{1}{2} \frac{\text{d} E_Z^\text{e}}{\text{d}B}$ and $E_Z^\text{e}(B)$ is the electron Zeeman splitting in Faraday geometry. 
By using the relation of Eq.~(\ref{eq:FWHM_Bex}), the electron exchange energy can be calculated from $\Delta B$ according to 
\begin{align}
E^\text{e}_\text{ex} = \frac{\gamma^\text{e}\Delta B}{2\sqrt{3}}.
\label{eq:FWHM_Eex}
\end{align}
For the studied QW  $\gamma^\text{e} = 1.6$~meV/T was measured via the giant Zeeman splitting effect in the Faraday geometry. With this value we get $E^\text{e}_\text{ex} =0.19$~meV.
%

\section{\label{sec:discussion}Discussion}

\begin{figure}[hbt]
\includegraphics[width=0.48\textwidth]{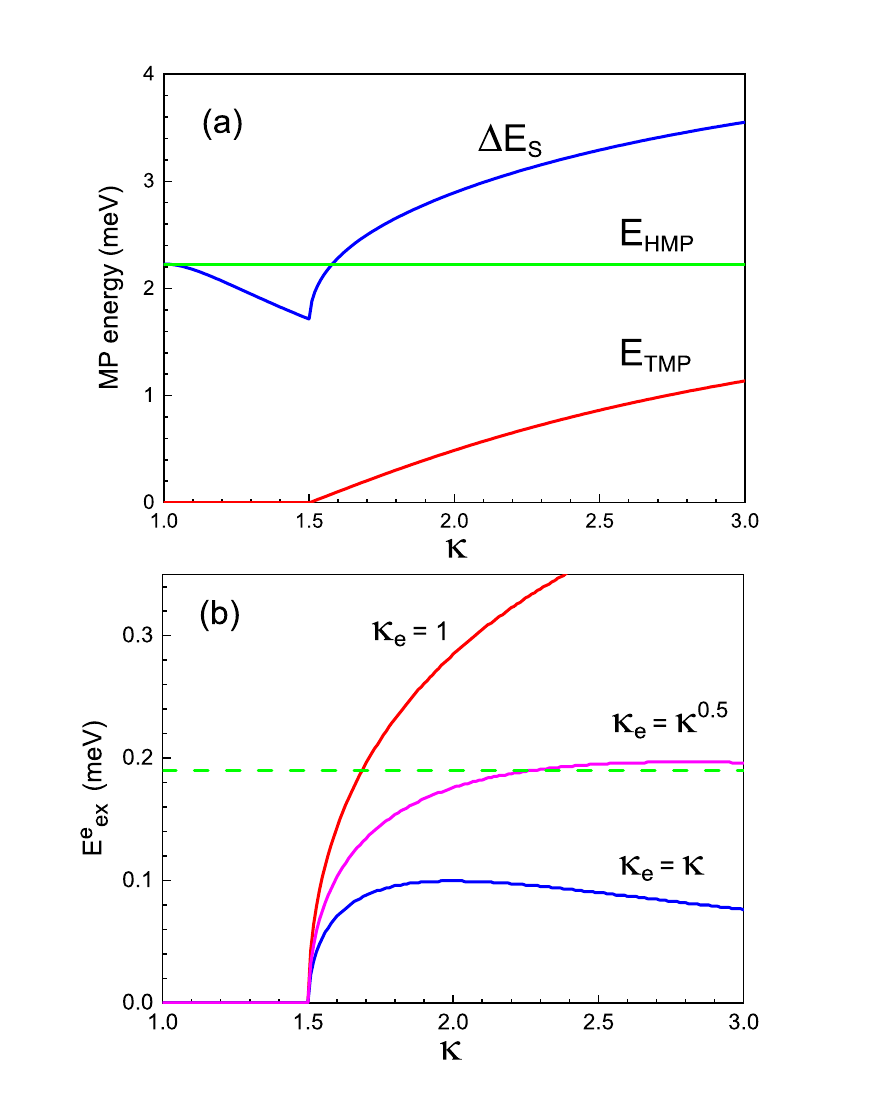}
\caption{ \label{modeling_Fig9} 
Energies characterizing magnetic polarons, calculated with the parameters $a_{\rm rh}=11$~nm and $\Delta_0=0.4$~meV. 
(a) T$^+$MP binding energy $E_{\rm TMP}$ (red line, Eq.~\eqref{eq:E_TMPkappa}), Stokes shift $\Delta E_{\rm S}$ of T$^+$MP (blue line, Eq.~\eqref{eq:Stokes_kappa}) and hole magnetic polaron binding energy $E_{\rm HMP}=2.2$~meV (green line, Eq.~\eqref{eq:E_HMPkappa}). 
(b) Electron exchange energy in T$^+$MP for various $\kappa_{\rm e}$. Green dashed line shows the value of 0.19~meV, evaluated from the experiment shown in Fig.~\ref{Polarization_Voigt}.
}
\end{figure}

In this section, we use the developed theory of T$^+$MP and perform model calculations for different polaron parameters and their dependence on $\kappa = a_\text{ph} / a_\text{rh}$, i.e., on the difference in localization sizes of the photoexcited and the resident holes. For the results presented in Fig.~\ref{modeling_Fig9}, we choose the parameters $a_{\rm rh}=11$~nm and $\Delta_0=0.4$~meV, which allow us to obtain reasonable agreement with the experimentally measured values and with the parameters known from literature, e.g., for the HMP in QWs with low Mn concentrations~\cite{zhukov_optical_2016,zhukov_optical_2019}. The binding energy of the T$^+$MP is shown by the red line in Fig.~\ref{modeling_Fig9}(a). It is obvious, that $E_{\rm TMP}=0$ for $\kappa = 1$, as in this case the exchange fields of the photoexcited and the resident holes exactly compensate each other ($a_\text{ph} = a_\text{rh}$). The T$^+$MP becomes stable for $\kappa >1.5$ and its binding energy monotonically increases up to 1~meV for $\kappa = 3$. As expected, the HMP binding energy  $E_{\rm HMP}=2.2$~meV shown by the green line is considerably larger than that of T$^+$MP. It is independent of $\kappa$ and depends only on $a_\text{rh}$, as only one resident hole  contributes to the HMP. The Stokes shift of T$^+$MP is shown by the blue line, and is about equal to the sum of the HMP and T$^+$MP binding energies: $\Delta E_{\rm S} \approx E_{\rm HMP} + E_{\rm TMP}$. 

Figure~\ref{modeling_Fig9}(b) shows calculations of the electron exchange energy in the developed T$^+$MP for different electron localization, controlled by the parameter $\kappa_\text{e}  = a_\text{e} / a_\text{rh}$. Three cases are shown: $\kappa_{\rm e}=\kappa$ (the electron radius equals to the radius of the photoexcited hole), $\kappa_{\rm e}=1$ (the electron radius equals to the radius of the resident hole), and $\kappa_{\rm e}=\sqrt{\kappa}$ (intermediate regime). The latter case corresponds to equal overlap integrals of the electron with the two holes. As shown in Section \ref{sec:model_Depol}, this relation ensures the absence of a residual PL polarization in strong transverse magnetic fields, as experimentally observed. 
The electron exchange energy of 0.19~meV, determined from the experimentally measured depolarization curve under selective excitation in Fig.~\ref{Polarization_Voigt}, is shown by the green dashed line. The reasonable agreement of the experiment and the modeling makes us convinced that the choice of parameters, $a_{\rm rh}=11$~nm and $\Delta_0=0.4$~meV, is a reasonable one. 

We have shown experimentally in Figs.~\ref{Exc_energy_dep_OO} and \ref{Polarization_Voigt} that the T$^+$MP properties under nonselective excitation differ considerably from those measured under selective excitation. The optical orientation degree is positive and reaches only 2\% which is a few times smaller than the negative OO under selective excitation. Also, the suppression by a magnetic field under nonselective excitation has a smaller FWHM of 0.2~T, corresponding to $E^{\rm e}_{\rm ex}=0.1$~meV, i.e. is twice smaller than in case of selective excitation. All these findings can be explained as follows: For nonselective excitation free excitons are photogenerated, excluding the mechanism responsible for the negative OO shown in Fig.~\ref{Trion_formation}. After their energy relaxation they are captured by HMPs formed by resident holes and together evolve into T$^+$MPs. The observed small $E^{\rm e}_{\rm ex}$ is, most likely, due to an on average larger electron localization radius for nonselective excitation.

\section{\label{sec:conclusion}Conclusions}

We have experimentally observed T$^+$MPs in (Cd,Mn)Te-based QWs and developed a model for their consideration. The T$^+$MP is a new type of two-dimensional magnetic polaron, which extends the family of previously reported EMPs and HMPs in DMS QWs. Similar to the EMP, the T$^+$MP has a finite lifetime limited by the trion recombination, but their formation processes differ considerably. The EMP formation starts from an unpolarized ensemble of Mn spins within the exciton localization volume, or to be more precise, from the small Mn polarization provided by thermal spin fluctuations of the Mn system \cite{yakovlev_magnetic_2010}. With EMP formation, the Mn polarization and the polaron binding energy increase. The formation process can be interrupted by exciton recombination and, therefore, in some cases, the observed polaron shift in experiment can be smaller than the equilibrium binding energy of the EMP.  

Contrary to that, the T$^+$MP formation process starts from a HMP involving a resident hole. The Mn polarization in the HMP is larger than that in the T$^+$MP. Therefore, the Mn polarization is reduced during the formation of the T$^+$MP and the Stokes shift under selective excitation is larger than the binding energy of the T$^+$MP. 

One may suggest that in QWs with resident electrons also a T$^-$MP can be observed, which is formed from a negatively charged exciton. Its formation would start from an electron MP, which due to the weak exchange field of the electron occurs mostly in the fluctuation regime~\cite{wolff_theory_1988}, i.e. the electron spin aligns along the magnetic fluctuations, but does not increase the Mn polarization. The binding energy of the T$^-$MP would be controlled by the hole exchange, which means that its formation process is provided by an increase of the Mn polarization, similar to the case of EMP formation. 

\begin{acknowledgments}

The authors acknowledge financial support by the Deutsche Forschungsgemeinschaft through the International Collaborative Research Centre TRR 160 (Project B4). F.G. and I.A.A. acknowledge financial support by the Deutsche Forschungsgemeinschaft (project No. AK40/11-1). R.R.A., B.R.N. and Yu.G.K. thank the Russian Foundation for Basic Research (Project No. 19-52-12066). K.V.K. acknowledges the Saint-Petersburg State University for the research Grant No. 91182694.

\end{acknowledgments}


\begin{thebibliography}{0}%
\makeatletter
\providecommand \@ifxundefined [1]{%
 \@ifx{#1\undefined}
}%
\providecommand \@ifnum [1]{%
 \ifnum #1\expandafter \@firstoftwo
 \else \expandafter \@secondoftwo
 \fi
}%
\providecommand \@ifx [1]{%
 \ifx #1\expandafter \@firstoftwo
 \else \expandafter \@secondoftwo
 \fi
}%
\providecommand \natexlab [1]{#1}%
\providecommand \enquote  [1]{``#1''}%
\providecommand \bibnamefont  [1]{#1}%
\providecommand \bibfnamefont [1]{#1}%
\providecommand \citenamefont [1]{#1}%
\providecommand \href@noop [0]{\@secondoftwo}%
\providecommand \href [0]{\begingroup \@sanitize@url \@href}%
\providecommand \@href[1]{\@@startlink{#1}\@@href}%
\providecommand \@@href[1]{\endgroup#1\@@endlink}%
\providecommand \@sanitize@url [0]{\catcode `\\12\catcode `\$12\catcode
  `\&12\catcode `\#12\catcode `\^12\catcode `\_12\catcode `\%12\relax}%
\providecommand \@@startlink[1]{}%
\providecommand \@@endlink[0]{}%
\providecommand \url  [0]{\begingroup\@sanitize@url \@url }%
\providecommand \@url [1]{\endgroup\@href {#1}{\urlprefix }}%
\providecommand \urlprefix  [0]{URL }%
\providecommand \Eprint [0]{\href }%
\providecommand \doibase [0]{https://doi.org/}%
\providecommand \selectlanguage [0]{\@gobble}%
\providecommand \bibinfo  [0]{\@secondoftwo}%
\providecommand \bibfield  [0]{\@secondoftwo}%
\providecommand \translation [1]{[#1]}%
\providecommand \BibitemOpen [0]{}%
\providecommand \bibitemStop [0]{}%
\providecommand \bibitemNoStop [0]{.\EOS\space}%
\providecommand \EOS [0]{\spacefactor3000\relax}%
\providecommand \BibitemShut  [1]{\csname bibitem#1\endcsname}%
\let\auto@bib@innerbib\@empty
\end{thebibliography}%


\begin{thebibliography}{}

\bibitem{awschalom_semiconductor_2002}
\emph{Semiconductor Spintronics and Quantum Computation}, eds. D. D. Awschalom, D. Loss and N. Samarth (Springer-Verlag, Berlin 2002).

\bibitem{zutic_spintronics_2004}
I. Žutić, J. Fabian and S. Das Sarma, Symmetry of anisotropic exchange interactions in semiconductor nanostructures, Rev. Mod. Phys. \textbf{76}, 323 (2004).

\bibitem{warnock_localized_1985}
J. Warnock, R. N. Kershaw, D. Ridgely, K. Dwight, A. Wold, and R. R. Galazka, Localized excitons and magnetic polaron formation in ({Cd},{Mn}){Se} and ({Cd},{Mn}){Te}, J. Luminescence \textbf{34}, 25 (1985).

\bibitem{dietl_dilute_2014}
T. Dietl and H. Ohno, Dilute ferromagnetic semiconductors: Physics and spintronic structures, Rev. Mod. Phys. \textbf{86}, 187 (2014).

\bibitem{crooker_terahertz_1996}
S. A. Crooker, J. J. Baumberg, F. Flack, N. Samarth, and D. D. Awschalom, Terahertz {Spin} {Precession} and {Coherent} {Transfer} of {Angular} {Momenta} in {Magnetic} {Quantum} {Wells}, Phys. Rev. Lett. \textbf{77}, 2814 (1996).

\bibitem{crooker_optical_1997}
S. A. Crooker, D. D. Awschalom, J. J. Baumberg, F. Flack, and N. Samarth, Optical spin resonance and transverse spin relaxation in magnetic semiconductor quantum wells, Phys. Rev. B \textbf{56}, 7574 (1997).

\bibitem{akimoto_larmor_1998}
R. Akimoto, K. Ando, F. Sasaki, S. Kobayashi, and T. Tani, {Larmor precession of {Mn}$^{2+}$ moments initiated by the exchange field of photoinjected carriers in {CdTe}/{Cd}$_{1-x}${Mn}$_x${Te} quantum wells}, Phys. Rev. B \textbf{57}, 7208 (1998).

\bibitem{wolff_theory_1988}
P. A. Wolff, Theory of {bound} {magnetic} {polarons} in {semimagnetic} {semiconductors}, in \emph{Semiconductors and {Semimetals}}, eds. J. K. Furdyna and J. Kossut (Academic Press, London 1988), pp. 413--454.

\bibitem{zhukov_optical_2019}
E. A. Zhukov, Yu. G. Kusrayev, E. Kirstein, A. Thomann, M. Salewski, N. V. Kozyrev, D. R. Yakovlev, and M. Bayer, Optical orientation of acceptor-bound hole magnetic polarons in bulk ({Cd},{Mn}){Te}, Phys. Rev. B \textbf{99}, 115204 (2019).

\bibitem{akimov_dynamics_2017}
I. A. Akimov, T. Godde, K. V. Kavokin, D. R. Yakovlev, I. I. Reshina, I. V. Sedova, S. V. Sorokin, S. V. Ivanov, Yu. G. Kusrayev and M. Bayer, Dynamics of exciton magnetic polarons in {CdMnSe}/{CdMgSe} quantum wells, Phys. Rev. B \textbf{95}, 155303 (2017).

\bibitem{seufert_dynamical_2001}
J. Seufert, G. Bacher, M. Scheibner, A. Forchel, S. Lee, M. Dobrowolska, and J. K. Furdyna, Dynamical {Spin} {Response} in {Semimagnetic} {Quantum} {Dots}, Phys. Rev. Lett. \textbf{88}, 027402 (2001).

\bibitem{zhukov_optical_2016}
E. A. Zhukov, Yu. G. Kusrayev, K. V. Kavokin, D. R. Yakovlev, J. Debus, A. Schwan, I. A. Akimov, G. Karczewski, T. Wojtowicz, J. Kossut, and M. Bayer, Optical orientation of hole magnetic polarons in ({Cd},{Mn}){Te}/({Cd},{Mn},{Mg}){Te} quantum wells, Phys. Rev. B \textbf{93}, 245305 (2016).

\bibitem{zhukov_spin_2007}
E. A. Zhukov, D. R. Yakovlev, M. Bayer, M. M. Glazov, E. L. Ivchenko, G. Karczewski, T. Wojtowicz, and J. Kossut, Spin coherence of a two-dimensional electron gas induced by resonant excitation of trions and excitons in {CdTe}/({Cd},{Mg}){Te} quantum wells, Phys. Rev. B \textbf{76}, 205310 (2007).

\bibitem{astakhov_exciton_2007}
G. V. Astakhov, A. V. Koudinov, K. V. Kavokin, I. S. Gagis, Yu. G. Kusrayev, W. Ossau, and L.W. Molenkamp, Exciton {Spin} {Decay} {Modified} by {Strong} {Electron}-{Hole} {Exchange} {Interaction}, Phys. Rev. Lett. \textbf{99}, 016601 (2007).

\bibitem{mackh_localized_1994}
G. Mackh, W. Ossau, D. R. Yakovlev, A. Waag, G. Landwehr, R. Hellmann, and E. O. G\"obel, Localized exciton magnetic polarons in CdMnTe, Phys. Rev. B \textbf{49}, 10248 (1994).

\bibitem{gaj_introduction_2010}
 \emph{Introduction to the {Physics} of {Diluted} {Magnetic} {Semiconductors}}, eds. J. A. Gaj and J. Kossut (Springer-Verlag, Berlin, Heidelberg 2010).

\bibitem{meier_optical_1984}
 \emph{Optical Orientation}, eds.  F. Meier and B. P. Zakharchenya (Elsevier, Amsterdam 1984).

\bibitem{merkulov_two-dimensional_1995}
I. A. Merkulov and K. V. Kavokin, Two-dimensional magnetic polarons, Phys. Rev. B \textbf{52}, 1751 (1995).

\bibitem{yakovlev_magnetic_2010}
D. R. Yakovlev and W. Ossau,  Magnetic polarons, in \emph{Introduction to the {Physics} of {Diluted} {Magnetic} {Semiconductors}}, eds. J. A. Gaj and J. Kossut (Springer-Verlag, Berlin, Heidelberg 2010) pp. 221--262.

\bibitem{kavokin_symmetry_2004}
K. V. Kavokin, Symmetry of anisotropic exchange interactions in semiconductor nanostructures, Phys. Rev. B \textbf{69}, 075302 (2004).  

\end{thebibliography}
\end{document}